\documentclass[letterpaper]{article} 
\usepackage{aaai24}  
\usepackage{times}  
\usepackage{helvet}  
\usepackage{courier}  
\usepackage[hyphens]{url}  
\usepackage{graphicx} 
\usepackage{natbib}  
\usepackage{caption} 

\usepackage{algorithm}
\usepackage{algorithmic}

\usepackage{microtype}
\usepackage{graphicx}
\usepackage{subcaption} 
\usepackage{booktabs} 
\usepackage{enumitem}

\usepackage{amsmath}
\usepackage{amssymb}
\usepackage{mathtools}
\usepackage{amsthm}

\usepackage[capitalize,noabbrev]{cleveref}

\usepackage{relsize} 
\DeclareMathOperator{\sign}{sgn}

\usepackage{multirow} 
\usepackage{comment}

\setlist[enumerate]{nolistsep}
\usepackage{color,soul}

\usepackage{booktabs}
\usepackage{longtable}
\usepackage{array}
\usepackage{multirow}
\usepackage{colortbl}
\usepackage{pdflscape}
\usepackage{threeparttable}
\usepackage{threeparttablex}
\usepackage{makecell}
\usepackage{xcolor}

\usepackage{newfloat}
\usepackage{listings}
\DeclareCaptionStyle{ruled}{labelfont=normalfont,labelsep=colon,strut=off} 
\floatstyle{ruled}
\newfloat{listing}{tb}{lst}{}
\floatname{listing}{Listing}
\title{On Feasibility of Intent Obfuscating Attacks}
\author{
    ZhaoBin Li\textsuperscript{\rm 1}, 
    Patrick Shafto\textsuperscript{\rm 1,\rm 2}
}
\affiliations{
    \textsuperscript{\rm 1}Department of Mathematics and Computer Science, Rutgers University--Newark, New Jersey, USA\\
    \textsuperscript{\rm 2}School of Mathematics, Institute for Advanced Study, New Jersey, USA\\
    zhaobin.li@rutgers.edu,
    patrick.shafto@rutgers.edu

}

\begin{document}

\maketitle

\begin{abstract}
Intent obfuscation is a common tactic in adversarial situations, enabling the attacker to both manipulate the target system and avoid culpability. Surprisingly, it has rarely been implemented in adversarial attacks on machine learning systems. We are the first to propose using intent obfuscation to generate adversarial examples for object detectors: by perturbing another non-overlapping object to disrupt the target object, the attacker hides their intended target. We conduct a randomized experiment on 5 prominent detectors---YOLOv3, SSD, RetinaNet, Faster R-CNN, and Cascade R-CNN---using both targeted and untargeted attacks and achieve success on all models and attacks. We analyze the success factors characterizing intent obfuscating attacks, including target object confidence and perturb object sizes. We then demonstrate that the attacker can exploit these success factors to increase success rates for all models and attacks. Finally, we discuss main takeaways and legal repercussions. If you are reading the AAAI/ACM version, please download the technical appendix on arXiv at \url{https://arxiv.org/abs/2408.02674} 
\end{abstract} 

\section{Introduction}

\begin{figure*}[ht!]
     \centering
     \begin{minipage}{0.5\textwidth}
         \centering
         \includegraphics[width=\textwidth]{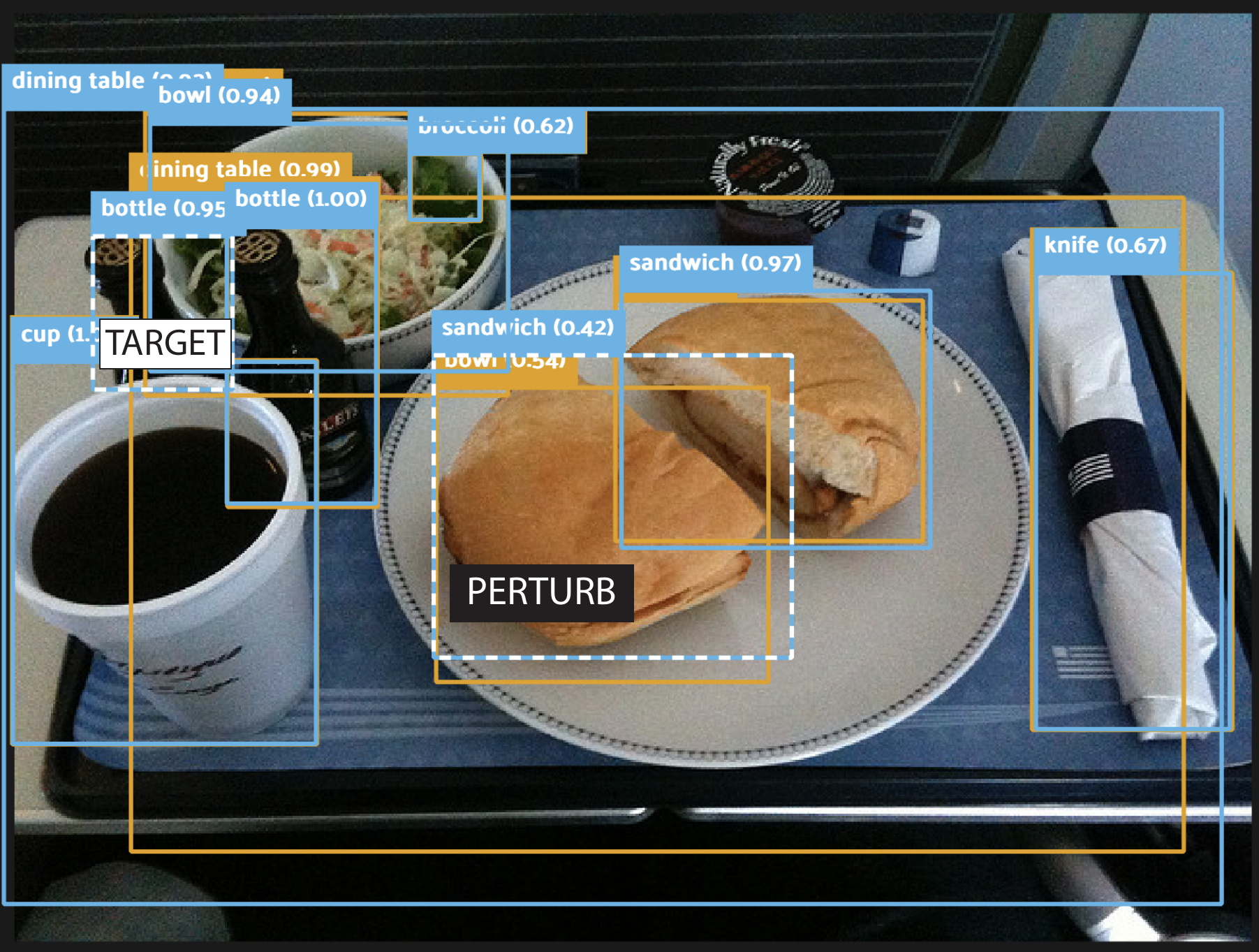}
         \caption{A vanishing attack perturbs a sandwich (dotted blue box) and causes YOLOv3 to miss the targeted bottle (no orange boxes are seen).}
         \label{fig:vanish_img}
         \vspace{0.02\textheight}
         \centering
         \includegraphics[width=\textwidth]{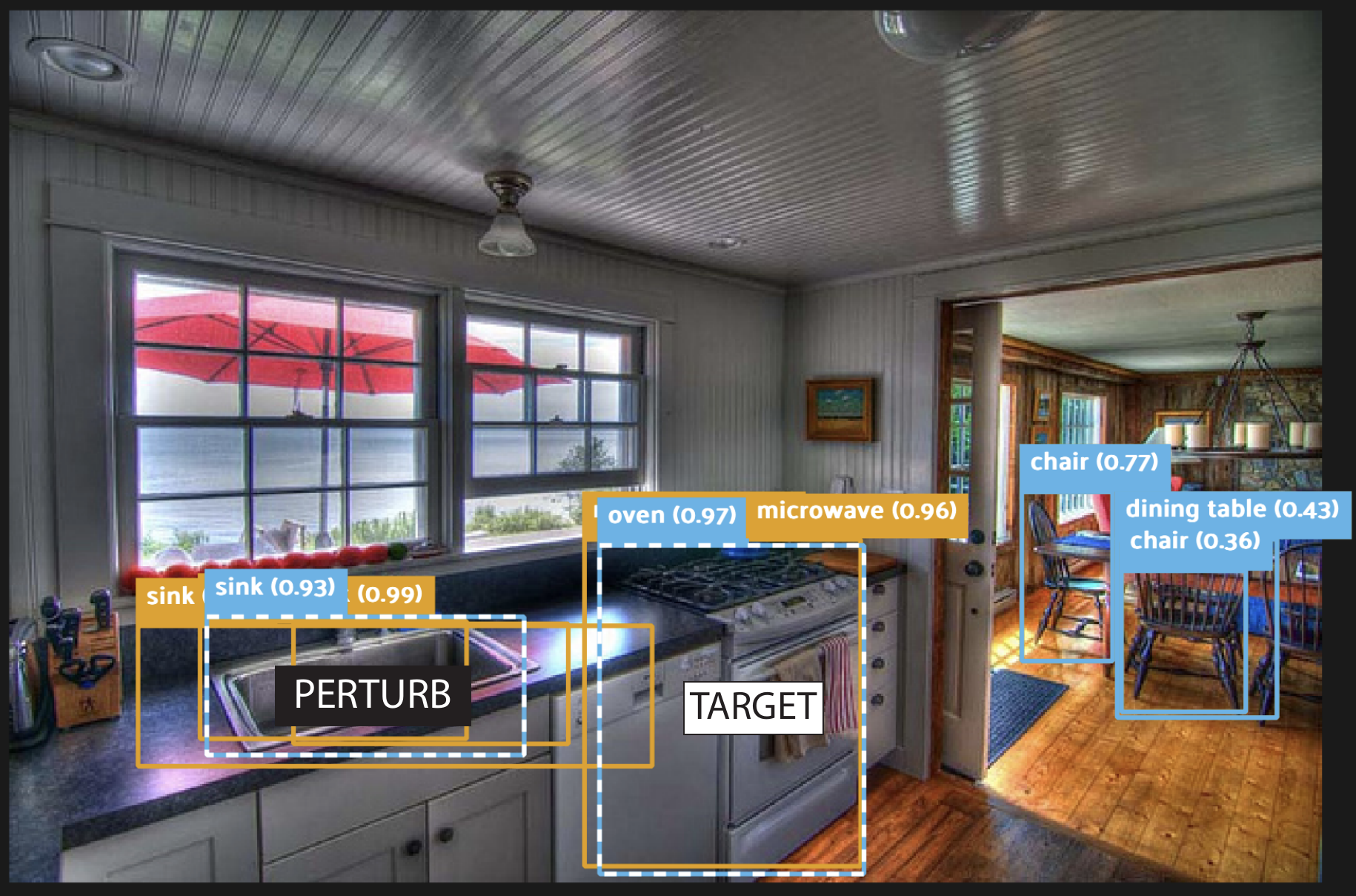}
         \caption{A mislabeling attack perturbs a sink and causes SSD to mislabel the targeted oven as a microwave with 0.96 confidence.}
         \label{fig:mislabel_img}
     \end{minipage}
     \hspace{0.05\textwidth}
     \begin{minipage}{0.4\textwidth}
         \centering
         \includegraphics[width=\textwidth]{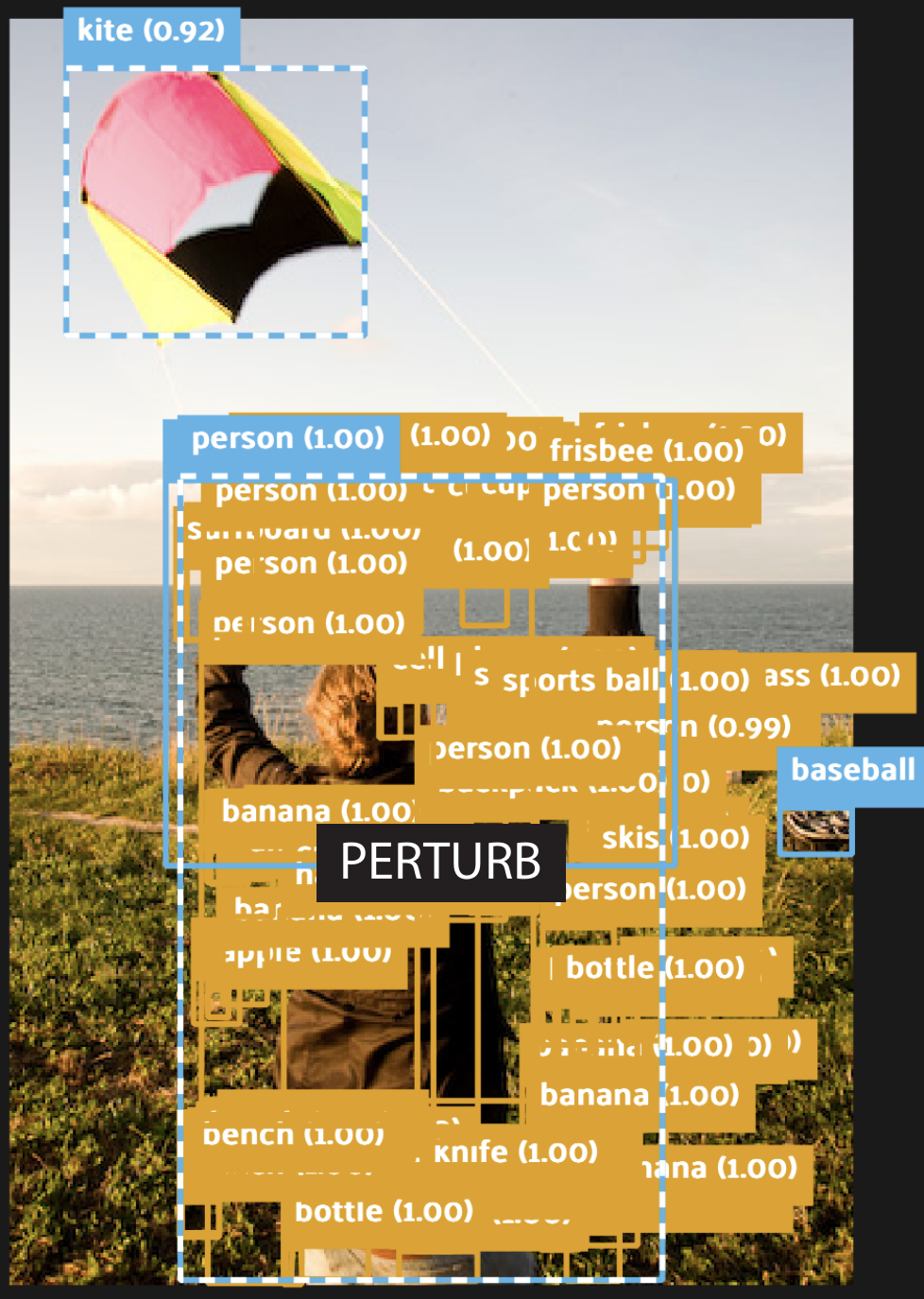}
         \caption{An untargeted attack perturbs a person and causes Faster R-CNN to miss the kite (and baseball) and hallucinate objects like bananas.}
         \label{fig:untargeted_img}
     \end{minipage}
        \caption{Disrupting a target object by perturbing another non-overlapping object enables intent obfuscating attacks to hide the attacker's intended target: The attacker can implement intent obfuscation using targeted (a) vanishing and (b) mislabeling attacks and (c) untargeted attacks, depending on their desired end result. Predictions on the original images are in blue and those on the adversarial images are in orange, with predictive confidence stated beside the class labels. The target and perturb objects are both dotted and labeled with ``target'' and ``perturb'' respectively. These examples are generated in the randomized experiment on the COCO dataset (Section \ref{sec:rand_att}). For clarity, the annotations are shown over the original images. Corresponding perturbed images are shown in Figure \ref{fig:rand_pert} in the appendix.}
        \label{fig:rand_img}
\end{figure*}

A malevolent agent sticks an adversarial patch to a bench on the sidewalk, causing a self-driving car to miss the stop sign and hit a crossing pedestrian. Upon interrogation, he claims no malicious intent; the patch is only an art. Because the sticker is on the bench but the effect is on the sign, authorities are unable to prove intent, preventing them from easily securing a conviction. This thought experiment highlights two serious implications of an intent obfuscating attack: it opens up new avenues for harmful exploits, and provides the culprit with ``plausible deniability''. 

Considering the potential significance of intent obfuscating attacks, it is important for the machine learning community to understand and defend against such attacks. Intent obfuscation, though a common practice in cyberattacks for penetrating target systems \citep{Lifars2020-ca}, has rarely been raised in the adversarial machine learning literature. Most research has focused on the competition between attack and defense, which involves crafting more effective adversarial examples to deceive machine learning systems and evade detection, and conversely more robust machine learning systems and more sensitive detection algorithms to mitigate attacks \citep{Ren2020-ws, Xu2020-wl}. Intent obfuscation complements the attack and defense literature by adding the dimension of intent to the competition: attackers can hide their purpose of attack for plausible deniability, and  defenders would have a harder time proving, or even determining, the purpose of attack from the adversarial examples.

We propose intent obfuscating attacks on object detectors through a contextual attack, in which we perturb one object to target another non-overlapping object. By attacking another object, intent is obfuscated providing plausible deniability, which conventional adversarial methods do not. As the opening example demonstrates, the attacker can manipulate an innocuous object to cause the detector to miss a critical target and simultaneously be legally shielded: they can blame the mistake on the machine learning system rather than admit to intentional deception. As a bonus, implementing intent obfuscation as a contextual attack opens up new avenues to attack the target, especially in situations where the attacker cannot manipulate the target directly. Moreover, contextual attacks are harder to detect since the defense algorithms not only need to inspect the target but also its surrounding region. The key question is whether perturbing one object to target another non-overlapping object is feasible on common detection models and object classes.

Feasibility is not guaranteed because object detectors are more complex than image classifiers. Detection involves both localization and classification, and its implementation varies widely across object detectors. The two most common types of object detectors \citep{Zhao2019-yz, Zou2019-ia} are 1 and 2-stage detectors. 2-stage detectors usually perform localization and then classification, whereas 1-stage detectors typically perform both tasks simultaneously. As a result, contextual attacks on object detectors are harder to implement and typically less general, since a method that succeeds on 1-stage detectors may not apply to 2-stage detectors. 
But intent obfuscating attacks could nevertheless achieve success by exploiting the contextual reasoning of object detectors---detectors are known to use contextual information to improve performance, either implicitly through end-to-end training \citep[e.g.\ YOLO][]{Redmon2015-rp} or explicitly through architectural design \citep[Section 2.4]{Tong2020-nl}. 

\begin{figure*}[tb]

{\centering \includegraphics[width=1\linewidth]{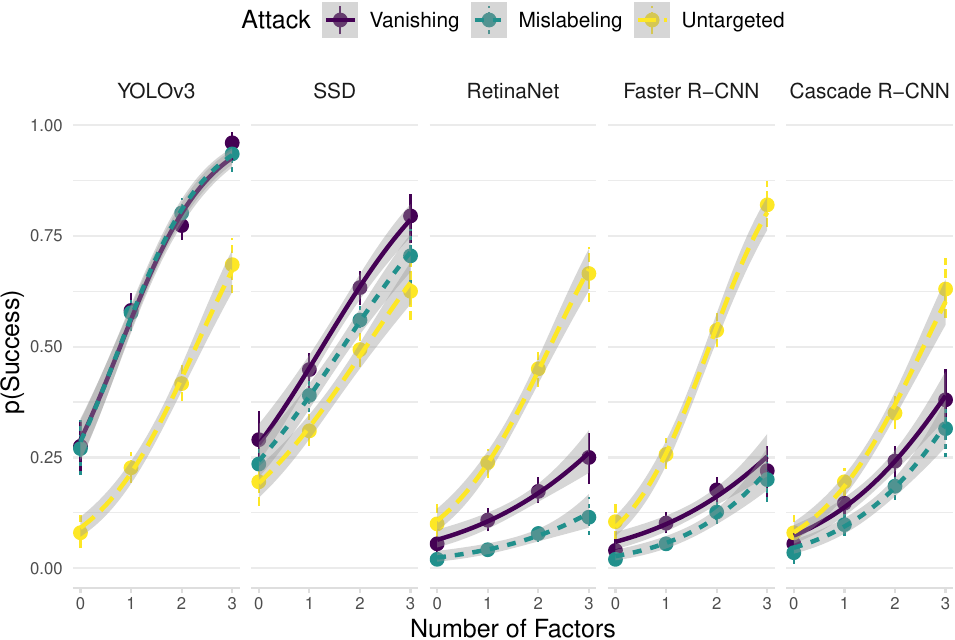} 

}

\caption{Success factors can be exploited in combination to significantly increase success rates:  We sampled target and perturb objects based on three validated success factors in Table \ref{tab:results_table} by targeting objects with low predicted confidence, perturbing large objects and selecting target and perturb objects close to one another. The binned summaries and regression trendlines graph success proportion against number of factors in the deliberate attack experiment. Errors are 95\% confidence intervals and every point aggregates success over 200 images. Success rates significantly increase as the number of factors combined increases. Significance is determined at $\alpha < 0.05$ using a Wald z-test on the logistic estimates. Full details are given in Section \ref{sec:del_per}.}\label{fig:biased_trend_graph}
\end{figure*}

\begin{figure*}[tb]

{\centering \includegraphics[width=1\linewidth]{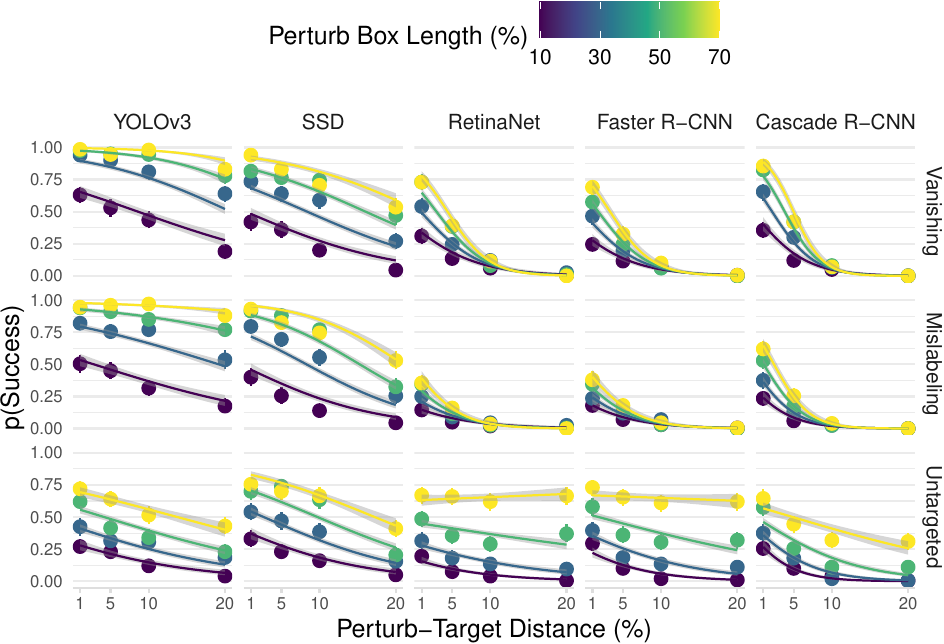} 

}

\caption{Perturbing an arbitrary region obfuscates intent with increased success for all models and attacks:  We implement intent obfuscating attack by perturbing an arbitrary non-overlapping square region to disrupt a randomly selected target object at various lengths and distances. The binned summaries and regression trendlines graph success proportion against perturb-target distance and perturb box length, both relative to image width or height, in the deliberate attack experiment. Errors are 95\% confidence intervals and every point aggregates success over 200 images. The deliberate attack multiplies success as compared to the randomized attack (Figure \ref{fig:success_trend_graph}), especially at close perturb-target distance and large perturb box length. Full details are given in Section \ref{sec:del_arb}.}\label{fig:arbitrary_trend_graph}
\end{figure*}

We implement intent obfuscating attacks on object detectors using the Targeted Objectness Gradient (TOG) algorithm \citep{Chow2020-ul} because TOG achieves greater success than previous attacks like DAG \citep{Xie2017-if}, according to \citet{Chow2020-rl}. In addition, as an iterative gradient-based algorithm, TOG can not only attack any modern state-of-the-art detector trained using backpropagation, but also enable the attacker to specify a precise target object for intent obfuscation. We apply TOG to both 1 and 2-stage detectors on the large-scale Microsoft Common Objects in Context (COCO) dataset \citep{Lin2014-ih}. We contribute to the important and understudied issue of intent obfuscation in adversarial machine learning:

\begin{enumerate}
    \item We are the first to propose an intent obfuscating attack on object detectors (Section \ref{sec:int_obf}).
    \item We determine the feasibility of intent obfuscating attacks on 5 prominent detectors---YOLOv3, SSD, RetinaNet, Faster R-CNN, and Cascade R-CNN---for both targeted and untargeted attacks (Section \ref{sec:rand_att}).
    \item We analyze the success factors for intent obfuscating attacks, including detection models, attack modes, target object confidence and perturb object sizes (Sections \ref{sec:rand_hp} and \ref{sec:rand_res}).
    \item We then exploit positive factors to increase success on all models and attacks by deliberately selecting perturb and target objects, as well as perturbing arbitrary regions, as shown in Figures \ref{fig:biased_trend_graph} and \ref{fig:arbitrary_trend_graph} respectively (Section \ref{sec:del_att}).
\end{enumerate}

\section{Related Work}

\textbf{Intent obfuscation:} Intent obfuscation is rare in the machine learning literature. One exception is a paper by \citet{Zhang2019-uy}, which investigates intent obfuscation in inverse reinforcement learning and applies the modeling results to an intrusion detection system. Another is a highly cited article on intent obfuscation by \citet{Sharif2016-jn}. The article uses adversarially patterned spectacles to conduct intent obfuscating attacks on face recognition systems and enable ``plausible deniability'' \citep[introduction]{Sharif2016-jn}. In comparison, we execute intent obfuscating attacks on object detectors, which is a more general and challenging problem. Moreover, as opposed to wearing conspicuously printed spectacles \citep[Figure 4 and 5]{Sharif2016-jn}, we use contextual attacks to obfuscate intent, which not only arouse less suspicion but also open up new avenues for manipulating the target.

\textbf{Contextual attacks:} Previous research has attempted to exploit the contextual reasoning of object detectors to improve existing attacks or to design new attacks \citep{Hu2021-oh, Saha2020-om, Lee2019-rz, Liu2018-pw, Zhang2020-uj, Cai2021-um}. The first 4 citations illustrate purely contextual attacks by perturbing non-overlapping regions, most notably through an adversarial patch. We extend those papers to cover greater breadth with 5 models, 3 attack modes and 80 COCO classes, as well as depth by systematically testing 10 success factors. More importantly, intent obfuscating attacks and contextual attacks diverge in 3 important aspects:

\begin{enumerate}
    \item \textbf{Aim:} Intent obfuscating attack aims to disrupt the target \textit{and} hide intent. Contextual attack is a means to obfuscate intent. Alternative means could include showing the detection system a manipulated image while recording the original image in the system logs.
    \item \textbf{Method:} Perturbing actual objects intuitively obfuscates intent more than perturbing a background region. A contextual attack does not distinguish the two.
    \item \textbf{Results:} We analyze success factors which preserve intent obfuscation through non-overlapping perturbations. For contextual attacks, an overriding factor for ensuring success is to perturb the target object together with its surrounding context, as shown in \cite{Zhang2020-uj}.
\end{enumerate}

\section{Intent Obfuscation}\label{sec:int_obf}

\subsection{Attack Methods}
We execute intent obfuscating attacks using the Targeted Objectness Gradient (TOG) algorithm \citep{Chow2020-ul}. TOG is an iterative gradient-based method similar to the Projected Gradient Descent (PGD) \citep{Madry2017-xe} attack and can be implemented both as untargeted and targeted attacks. We are most interested in the targeted attack because it gives the attacker precise control over the desired end result. A targeted attack achieves its purpose by manipulating the ground-truth for training the object detector. \footnote{For object detection, the ground-truth for a labeled object comprise 4 bounding box coordinates and 1 class label.} The attacker can aim for the detector to mislabel the target object by changing its class label and retaining its original bounding box (``mislabeling'' attack), or for the target object to vanish entirely by removing both its bounding box and class label from the ground-truth (``vanishing'' attack). Their technical details are elaborated below: 

Let $\theta$ be the model parameters, $x$ the input image, $y'$ the desired target, and $L(\theta, x, y')$ the optimization loss. The desired target $y'$ could be derived by manipulating either the ground-truth or the model predictions. At iteration $t + 1$, we add the signed gradients $\nabla_x L(\theta, x, y')$ times the learning rate $\alpha$ to the perturbed image in the previous iteration $x^t$. Then we limit the change in $x$ to within the bounds $S$ and iterate the process for a total of $T$ iterations:

\begin{equation}\label{eq:targeted}
x^{t+1} = \mathlarger{\Pi}_{x + S} \left[x^t - \alpha \cdot \sign(\nabla_x L(\theta, x, y'))\right]
\end{equation}

Whereas a targeted attack minimizes the training loss towards the desired target, an untargeted attack maximizes (note the change in sign) the training loss $L(\theta, x, y)$ towards the original target $y$, which could either be the ground-truth or the model predictions:

\begin{equation}\label{eq:untargeted}
x^{t+1} = \mathlarger{\Pi}_{x + S} \left[x^t + \alpha \cdot \sign(\nabla_x L(\theta, x, y))\right]
\end{equation}

The optimization loss $L$ depends on the model, which we will present in the next section. Since the attacker will not have access to the ground-truth in most scenarios, we will conduct experiments by using the model predictions as $y$. 

\subsection{Model Losses}\label{app:mod_los}
We attack 5 prominent detection models---comprising 3 1-stage detectors (SSD, YOLOv3, and RetinaNet) and 2 2-stage detectors (Faster R-CNN and Cascade R-CNN)---implemented in the versatile MMDetection toolbox \citep{Chen2019-nt} and pretrained on the COCO dataset \citep{Lin2014-ih}. All models, besides the more recent and highly cited Cascade R-CNN, are spotlighted in reviews by \citet{Zhao2019-yz} and \citet{Zou2019-ia} and stated as the most widely implemented according to \citet{Papers_With_Code2024-ju}. Table \ref{tab:models_table} summarizes the 5 detection models and corresponding attack losses. Full details are given below:

\textbf{YOLOv3:} YOLOv3 \citep{Redmon2018-ud} prioritizes speed and uses a single convolutional network to predict bounding boxes and class labels. The class label is described by the objectness score, defined as the probability that the bounding box contains an object, and the class probability conditioned on the objectness score. Consequently, YOLOv3 has 3 training losses: the objectness loss, the class loss and the box regression loss \citep[equation 3]{Redmon2015-rp}. We attack the objectness loss for the vanishing attack and the class loss for the mislabeling attack. For untargeted attack, we attack all training losses. Additionally, YOLOv3 is optimized through end-to-end training and ``implicitly encodes contextual information'' \citep[introduction]{Redmon2015-rp}. Therefore, it should be more vulnerable to contextual attacks. In the experiment, we use a pretrained YOLOv3 with a DarkNet-53 backbone and input size 608 $\times$ 608. The model achieves 33.7 COCO mean average precision (mAP), the primary metric in the COCO challenge \citep{Coco2024-tc}.

\textbf{SSD:} Like YOLOv3, SSD \citep{Liu2015-lo} also uses a single convolutional network and is optimized through end-to-end training, improving both speed and accuracy. Uniquely, SSD adds several convolutional layers which successively decrease in sizes after the base network. These layers predict bounding boxes at multiple sizes and aspect ratios. The training losses in SSD include box regression loss and class loss. Since the class loss includes the background class in addition to the 80 COCO class labels, we target the class loss for both vanishing and mislabeling attacks. For untargeted attack, we attack all training losses. In the experiment, we use a pretrained SSD with a VGG-16 backbone \citep{Simonyan2014-vo} and input size 512 $\times$ 512. The model achieves 29.5 COCO mAP.

\textbf{RetinaNet:} RetinaNet \citep{Lin2017-va} uses a novel Focal Loss to address class imbalance in training 1-stage detectors: most training examples belong to the easily categorized background class and thereby overwhelm the training signal. Focal Loss mitigates the issue by down-weighting easily categorized background examples during training to emphasize the harder object examples and thereby increases training accuracy. RetinaNet also incorporates convolutional layers structured as a Feature Pyramidal Network (FPN) \citep{Lin2017-mx} for multi-scale detection. Like SSD, RetinaNet's training losses comprise both the class loss (which includes the background class) and bounding box loss. We target the class loss for both vanishing and mislabeling attacks. For untargeted attack, we attack all training losses. In the experiment, we use a pretrained RetinaNet with a ResNet-50 backbone \citep{He2015-we}. The model achieves 36.5 COCO mAP.

\textbf{Faster R-CNN:} Faster R-CNN \citep{Ren2015-hq} adds a region proposal network (RPN) to the detection network in Fast R-CNN \citep{Girshick2015-vv} to improve both speed and accuracy. Faster R-CNN begins detection with a base network to extract convolutional features. Then using these convolutional features, the RPN proposes object regions with associated objectness scores. The detection network then uses both the convolutional features and region proposals to predict bounding boxes and class labels. Hence, Faster R-CNN has 4 training losses: the box regression loss and objectness loss in the RPN and the box regression loss and class loss in the detection network. Since the class loss for the detection network also includes the background class in addition to the 80 COCO class labels \citep[equation 1]{Girshick2015-vv}, we attack both the class loss and objectness loss for the vanishing attack and attack only the class loss for the mislabeling attack. For untargeted attack, we attack all training losses. In the experiment, we use the pretrained Faster R-CNN with a ResNet-50 backbone and FPN. The model achieves 37.4 COCO mAP.

\textbf{Cascade R-CNN:} Cascade R-CNN \citep{Cai2017-wp} extends the Faster R-CNN architecture with a cascade structure to generate more accurate detections. Cascade R-CNN repeats the RPN stage in Faster R-CNN thrice to increase proposals quality. The 2nd and 3rd RPNs in Cascade R-CNN also propose class labels (which include the background class) rather than only the objectness score in the 1st RPN. All 3 RPNs also predict bounding box coordinates. Hence, the training losses for Cascade R-CNN comprise 4 box regression losses, 3 class losses and 1 objectness loss. We attack the objectness loss and class losses for the vanishing attack and attack all class losses for the mislabeling attack. For untargeted attack, we attack all training losses. In the experiment, we use a pretrained Cascade R-CNN with a ResNet-50 backbone and FPN. The model achieves 40.3 COCO mAP.

\begin{table*}[htb]
\centering
\begin{threeparttable}
\begin{tabular}[t]{llllll}
\toprule
\multicolumn{3}{c}{ } & \multicolumn{3}{c}{Attack Losses\textsuperscript{c}} \\
\cmidrule(l{3pt}r{3pt}){4-6}
\multicolumn{3}{c}{ } & \multicolumn{2}{c}{Targeted} & \multicolumn{1}{c}{ } \\
\cmidrule(l{3pt}r{3pt}){4-5}
Detectors & Stages\textsuperscript{a} & COCO mAP\textsuperscript{b} & Vanishing & Mislabeling & Untargeted\textsuperscript{d}\\
\midrule
\cellcolor{gray!10}{YOLOv3} & \cellcolor{gray!10}{1} & \cellcolor{gray!10}{33.7} & \cellcolor{gray!10}{Object} & \cellcolor{gray!10}{Class} & \cellcolor{gray!10}{Class, Box, Object}\\
SSD & 1 & 29.5 & Class & Class & Class, Box\\
\cellcolor{gray!10}{RetinaNet} & \cellcolor{gray!10}{1} & \cellcolor{gray!10}{36.5} & \cellcolor{gray!10}{Class} & \cellcolor{gray!10}{Class} & \cellcolor{gray!10}{Class, Box}\\
\makecell[l]{Faster\\R-CNN} & 2 & 37.4 & \makecell[l]{RPN: Object;\\Det: Class} & Det: Class & \makecell[l]{RPN: Object, Box;\\Det: Class, Box}\\
\cellcolor{gray!10}{\makecell[l]{Cascade\\R-CNN}} & \cellcolor{gray!10}{2} & \cellcolor{gray!10}{40.3} & \cellcolor{gray!10}{\makecell[l]{RPN 1: Object;\\RPNs 2, 3 + Det: Class}} & \cellcolor{gray!10}{\makecell[l]{RPNs 2, 3: Class;\\Det: Class}} & \cellcolor{gray!10}{\makecell[l]{RPN 1: Object, Box;\\RPNs 2, 3 + Det: Class, Box}}\\
\bottomrule
\end{tabular}
\begin{tablenotes}
\item[a] In general, 1-stage detectors are quicker whereas 2-stage detectors are more accurate, though the 1-stage RetinaNet aims to be both quick and accurate. In a 2-stage detector, the input image passes through a Region Proposal Network (RPN) stage and a detection (Det) stage.
\item[b] COCO mean Average Precision (mAP) is the primary metric on the COCO challenge.
\item[c] The training losses in detectors typically include the box regression loss (Box), the class loss on the 80 COCO labels and/or the background class (Class), and the objectness loss on categorizing an image region as background or object (Object).
\item[d] Untargeted attack targets all training losses in a model, i.e. the backpropagation loss.
\end{tablenotes}
\end{threeparttable}
\centering
\caption{\label{tab:models_table}Detection models and attack losses. Full details are given in Appendix \ref{app:mod_los}.}
\end{table*}

\section{Randomized Attack}\label{sec:rand_att}

\subsection{Setup}\label{sec:rand_set}
We evaluate the 3 intent obfuscating attacks---vanishing, mislabeling and untargeted---on the 5 models using the 2017 COCO dataset \citep{Lin2014-ih}. The COCO dataset has 80 categories of common objects in everyday scenes for object detection and the 2017 split has 118,000 train images and 5,000 test images \citep{Papers_with_Code2024-ha}. We use the test images to attack the 5 models with pretrained weights obtained through MMDetection \citep{Chen2019-nt} and visualized the results using the FiftyOne visualization app \citep{Moore_B_E_and_Corso_J_J2020-xp}. 

\textbf{Target and perturb objects selection:} First, we evaluate the models on the original images and count a detection as correct when both the bounding box and the class label match the ground-truth with at least 0.3 intersection-over-union (IOU) and 0.3 confidence respectively. Note that we do not use the standard COCO mean average precision (mAP) metric since mAP measures detection precision over the whole dataset, but we are interested in evaluating success for single objects. After getting the initial predictions, we restrict only to the correctly predicted objects. Then we randomly sample a target object and another \textit{non-overlapping} perturb object per image. Images with less than 2 correctly predicted non-overlapping objects are ignored. 

\textbf{Ground-truth manipulation for targeted attack:} Then we create the desired target $y'$ from the ground-truth $y$ for the 2 targeted attacks (vanishing and mislabeling equation \ref{eq:targeted}). For the vanishing attack, we remove the target object entirely---both the class label and bounding box---from the ground-truth $y$ to get $y'$. For the mislabeling attack, we change the class label of the target object in $y$ to a random class (``intended class'' from now on) to get the desired target $y'$. For the untargeted attack, we evaluate the randomly selected target object only to compare success rates with the 2 targeted attacks. 

\textbf{Attack parameters:} Next, we run the 3 attacks using iterations 10, 50, 100, and 200, but not more than 200 since success rates plateau after. For every iteration, we set a learning rate $\alpha$ which could maximally change a pixel from 0 (black) to 1 (white). For instance, we use a 0.1 learning rate for 10 iterations. In addition, we set a perturbation bound $S$ such that the image remains in its original range of $[0, 1]$ after every iteration. 
We also repeated the simulations with an $l_{\infty}$-norm of 0.05 applied after every iteration. Since the norm constraint is not central to intent obfuscating attacks, we put its results in the appendix. For every model, attack and iteration combination, we resampled 4,000 test images.

\textbf{Results evaluation:} We distort the bounding box of the perturb object and then re-evaluate the generated adversarial image: as in the initial evaluation step, we use IOU and confidence thresholds of 0.3 to determine whether the attack succeeds in disrupting the target object. The attack speed mainly depended on model complexity. More experimental details are included in Appendix \ref{app:rand_set}.

\subsection{Hypotheses}\label{sec:rand_hp}

We conducted a thorough analysis by listing 10 hypotheses increasing success rates and systematically testing whether those hypotheses are valid in the next section. For all attacks, we expect to achieve higher success rates for:
\begin{enumerate}
    \item \label{hp:model_stage} \textbf{1-stage (YOLOv3, SSD, and RetinaNet) than 2-stage (Faster R-CNN and Cascade R-CNN) detectors:} intuitively, perturbing an input pixel to change one loss component in an intended direction is easier than for multiple loss components. As the number of loss components increases, the chances that the same perturbation will change all losses in the same direction decreases, making the overall attack harder. Because we attack more loss components for 2-stage than 1-stage detectors, we expect to achieve correspondingly lower success rates for 2-stage detectors, beyond what could be explained by their higher COCO mAPs listed in Table \ref{tab:models_table}.
    \item \label{hp:target_untarget} \textbf{Targeted than untargeted attack:} the gradient signal in a targeted attack is precisely aimed at the target object, whereas for an untargeted attack the gradient signal is broadly aimed at all objects in the image. Therefore, the chances that an untargeted attack disrupts the target object is lower.
    \item \label{hp:vanish_mislabel} \textbf{Vanishing than mislabeling attack:} converting the original class label to the background class should be easier than to non-background classes, since the background class contains everything not labeled in the COCO dataset and thereby makes up a large portion of the input space.
    \item \label{hp:num_iteration} \textbf{Larger attack iterations:} we expect larger attack iterations to achieve better local minima and maxima respectively for targeted and untargeted attacks since more iterations allow more possible routes to navigate across the loss landscape.
    \item \label{hp:target_conf} \textbf{Target objects with lower predicted confidence:} the higher the predicted confidence, the larger the decrease in class probability needed to achieve success and the more the attack has to perturb the class loss.
    \item \label{hp:perturb_bbox} \textbf{Perturb objects with larger bounding boxes:} larger bounding boxes enable the attack to perturb more pixels, after controlling for Hypothesis \ref{hp:object_dist}.
    \item \label{hp:object_dist} \textbf{Shorter distance between perturb and target objects:} since object detectors likely utilize nearby context to make predictions, perturbing nearby pixels should change the predictions more. Because larger perturb objects (Hypothesis \ref{hp:perturb_bbox}) are more likely to be closer to the target object, we will control for both with a regression model.
    \item \label{hp:target_success} \textbf{Target object classes with lower COCO mean accuracy:} when an object detector achieves lower mean accuracy for particular classes on the COCO dataset, attacking target objects belonging to those classes should be easier. When the target object class has lower mean accuracy, the target object will likely be predicted with lower confidence. Considering Hypothesis \ref{hp:target_conf}, we will also control for the latter.
\end{enumerate}

For specific attacks, we expect to achieve higher success rates for 
\begin{enumerate}
    \setcounter{enumi}{8}
    \item \label{hp:untarget_iou} \textbf{Target objects with lower intersection-over-union (IOU) for the untargeted attack:} the lower the IOU of predicted and ground-truth bounding boxes, the less the untargeted attack has to perturb the box loss to misalign the detection to less than the IOU threshold.
    \item \label{hp:mislabel_conf} \textbf{Intended classes with higher probabilities for the mislabeling attack:} in a mislabeling attack we aim to change the target prediction to the intended class. When the intended class has higher probability on the original image, the increase in probability of the intended class required for the detector to mislabel the target is smaller, and the attack would have to change the class loss by a lesser degree. The reasoning is similar to the one in Hypothesis \ref{hp:target_conf}. In addition, since higher probability of the intended class likely entails lower confidence of the predicted class \footnote{To be clear, class probability and confidence are the same. In alignment with the object detection literature, I will use confidence to mean probability only for the predicted class.}, we will also control for the latter.
\end{enumerate}

\subsection{Results}\label{sec:rand_res}
\begin{figure*}[tb]

{\centering \includegraphics[width=1\linewidth]{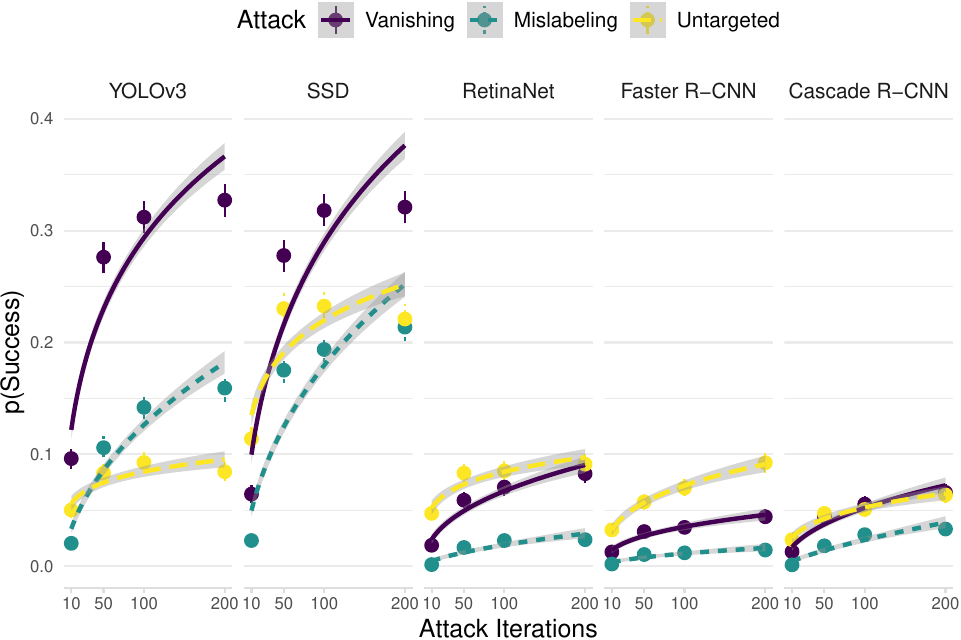} 

}

\caption{Intent obfuscating attack is feasible for all models and attacks:  We conduct a randomized experiment by resampling COCO images, and within those images randomly sampling correctly predicted target and perturb objects. Then we distort the perturb objects to disrupt the target objects varying the attack iterations. The binned summaries and regression trendlines graph success proportion against attack iterations in the randomized attack experiment. Errors are 95\% confidence intervals and every point aggregates success over 4,000 images. Targeted vanishing and mislabeling attacks obtain significantly greater success on the 1-stage YOLOv3 and SSD than the 2-stage Faster R-CNN and Cascade R-CNN detectors. However, the 1-stage RetinaNet is as resilient as the 2-stage detectors. Moreover, success rates significantly increase with larger attack iterations. Significance is determined at $\alpha < 0.05$ using a Wald z-test on the logistic estimates. Full details are given in Section \ref{sec:rand_att}.}\label{fig:success_trend_graph}
\end{figure*}

The success rates without norm constraint are shown in Figure \ref{fig:success_trend_graph}. Imposing a 0.05 $l_{\infty}$-norm constraint slightly decreases success, as shown in Figure \ref{fig:success_trend_graph_normed} in the appendix, but the trends remain the same. Hence, we will only conduct hypothesis testing on the results without norm constraint.  

For all hypotheses, we use logistic regression to determine if the stated variables significantly predict success rates. We transform the predictors as appropriate and run separate regressions for every model and attack combination, unless the predictor variable includes model (Hypothesis \ref{hp:model_stage}) or attack (Hypotheses \ref{hp:target_untarget} and \ref{hp:vanish_mislabel}). Except for testing the effect of iterations (Hypothesis \ref{hp:num_iteration}), we restrict the data to the maximum 200 attack iterations to analyze the strongest possible results. We computed the p-values using a Wald z-test and set the significance level ($\alpha$) to the usual 0.05. Attacked images are illustrated in Figure \ref{fig:rand_img} and hypotheses and results are summarized in Table \ref{tab:results_table}. We will state the conclusions below. Graphs and tabulated  statistics are in Appendix \ref{app:rand_tab}.

\begin{enumerate}
    \item \label{res:model_stage} \textbf{1-stage (YOLOv3, SSD, and RetinaNet) than 2-stage (Faster R-CNN and Cascade R-CNN) detectors:} As shown in Figure \ref{fig:success_trend_graph}, both vanishing and mislabeling attacks achieve significantly higher success rates for 1-stage than 2-stage detectors. The higher success on 1-stage detectors could not be explained by their lower COCO mAPs. Surprisingly, the 1-stage RetinaNet is as robust as 2-stage detectors---training RetinaNet using Focal Loss not only boosts COCO accuracy but also increases resilience against intent obfuscating attacks (Table \ref{tab:model_stage_table}). 
    \item \label{res:target_untarget} 
    \textbf{Targeted than untargeted attack:} The results are mixed: targeted attack is significantly more successful than untargeted attack for YOLOv3 and slightly more successful for SSD, but the increase is non-existent or reversed for RetinaNet, Faster R-CNN and Cascade R-CNN (Table \ref{tab:target_untarget_vanish_mislabel_table} and Figure \ref{fig:success_trend_graph}). As stated in Result \ref{res:model_stage}, RetinaNet, Faster R-CNN and Cascade R-CNN are more robust than YOLOv3 and SSD against intent obfuscating attack, and perhaps more robust models require a coordinated attack against all loss components to achieve success.     
    \item \label{res:vanish_mislabel} \textbf{Vanishing than mislabeling attack:} Vanishing attack achieves significantly more success than mislabeling attack for all models (Table \ref{tab:target_untarget_vanish_mislabel_table} and Figure \ref{fig:success_trend_graph}).
    \item \label{res:num_iteration} \textbf{Larger attack iterations:} Larger attack iterations (log-transformed) significantly increase success for all models and attacks (Table \ref{tab:num_iteration_table}). 
    \item \label{res:target_conf} \textbf{Target objects with lower predicted confidence:} Lower target confidence significantly increases success rates for all models and attacks (Table \ref{tab:target_conf_table} and Figure \ref{fig:target_conf_graph}).
    \item \label{res:perturb_bbox} \textbf{Perturb objects with larger bounding boxes:} Larger perturb objects significantly increase success rates for all models and attacks, except for mislabeling attacks on Faster R-CNN, after controlling for perturb-target distances (Table \ref{tab:perturb_bbox_and_object_dist_table} and Figure \ref{fig:perturb_bbox_and_object_dist_graph}).
    \item \label{res:object_dist} \textbf{Shorter distance between perturb and target objects:} Shorter perturb-target distances significantly increase success rates for all models and attacks, after controlling for perturb object sizes (Table \ref{tab:perturb_bbox_and_object_dist_table} and Figure \ref{fig:perturb_bbox_and_object_dist_graph}).
    \item \label{res:target_success} \textbf{Target classes with lower COCO mean accuracy:} The results are mixed: of the 15 model and attack combinations, higher COCO class accuracy significantly decreases success rates for 5 combinations but increases success rates for 4, after controlling for target class confidence. The relatively large interaction terms make interpretation challenging (Table \ref{tab:target_success_table} and Figure \ref{fig:target_success_graph}).
    \item \label{res:untarget_iou} \textbf{Target objects with lower intersection-over-union (IOU) for the untargeted attack:} Lower IOU increases success rates for untargeted attack on all models (Table \ref{tab:untarget_iou_table} and Figure \ref{fig:untarget_iou_graph}). 
    \item \label{res:mislabel_conf} \textbf{Intended classes with higher probabilities for the mislabeling attack:} The results are mixed: higher intended class probability (log-transformed) does \textit{not} predict success rates for mislabeling attack after controlling for target class confidence for SSD, Faster R-CNN, and Cascade R-CNN. However, it is significantly negative for YOLOv3 and positive for RetinaNet. (Table \ref{tab:mislabel_conf_table} and Figure \ref{fig:mislabel_conf_graph}). 
\end{enumerate}



\begin{table}[htb]
\centering
\begin{tabular}[t]{>{\raggedright\arraybackslash}p{1.25in}>{\raggedright\arraybackslash}p{1.5in}}
\toprule
Hypotheses (higher success for) & Accepted (across attacks and models)\textsuperscript{a}\\
\midrule
\cellcolor{gray!10}{1-stage $>$ 2-stage models (YOLOv3, SSD, RetinaNet $>$ Faster R-CNN, Cascade R-CNN)} & \cellcolor{gray!10}{YOLOv3, SSD $>$ RetinaNet, Faster R-CNN, Cascade R-CNN in vanishing and mislabeling attacks (1-stage RetinaNet is as resilient as 2-stage models)}\\
Targeted $>$ Untargeted attack & YOLOv3 only\\
\cellcolor{gray!10}{Vanishing $>$ Mislabeling attack} & \cellcolor{gray!10}{All}\\
Larger attack iterations & All\\
\cellcolor{gray!10}{Less confident targets} & \cellcolor{gray!10}{All}\\
\addlinespace
Larger perturb boxes & All except mislabeling attack on Faster R-CNN\\
\cellcolor{gray!10}{Shorter perturb-target distance} & \cellcolor{gray!10}{All}\\
Less accurate target COCO class & Mixed\\
\cellcolor{gray!10}{Lower target IOU\textsuperscript{b} (untargeted attack only)} & \cellcolor{gray!10}{All}\\
More probable intended class (mislabeling attack only) & Mixed\\
\bottomrule
\multicolumn{2}{l}{\rule{0pt}{1em}\textsuperscript{a} $p < .05$ for Wald z-test on logistic estimate}\\
\multicolumn{2}{l}{\rule{0pt}{1em}\textsuperscript{b} intersection-over-union}\\
\end{tabular}
\centering
\caption{\label{tab:results_table}Hypothesis testing in the randomized attack (Sections \ref{sec:rand_hp} and \ref{sec:rand_res})}
\end{table}

\section{Deliberate Attack}\label{sec:del_att}

Rather than randomly selecting target and perturb objects in the randomized experiment, the attacker can---and will---select objects to exploit the success factors listed in the previous section. For instance, to maximize havoc on a congested street, he may target the stop sign with the lowest predicted confidence (Result \ref{res:target_conf}) and use a vanishing attack if most self-driving cars use a detector based on YOLO (Result \ref{res:model_stage}). He could also increase success by deliberately perturbing larger objects (Result \ref{res:perturb_bbox}) closer to the target (Result \ref{res:object_dist}). Moreover, he can easily multiply success on a random target for any detector by perturbing a large \textit{arbitrary} region close to the target object. 
We run experiments for the two common scenarios of deliberately selecting target and perturb objects and perturbing an arbitrary region in Sections \ref{sec:del_per} and \ref{sec:del_arb} respectively.

\subsection{Selecting Easier Targets}\label{sec:del_per}
Building on our randomized attacks described in Section \ref{sec:rand_att}, we deliberately exploit 3 validated success factors in Table \ref{tab:results_table} to select:

\begin{enumerate}
    \item Target objects with less than 0.5 predicted confidence.
    \item Perturb objects with bounding boxes more than 25\% of the image size.
    \item Perturb and target objects with distances less than 25\% across the image. \footnote{We use an algorithm in game development \citep{Congusbongus2018-yv} to compute the minimum distances between the perturb and target bounding boxes. We set the image width and height to 1 and select perturb and target objects with distances less than 0.25.}
\end{enumerate}

\begin{figure}[htb]

{\centering \includegraphics[width=0.99\columnwidth]{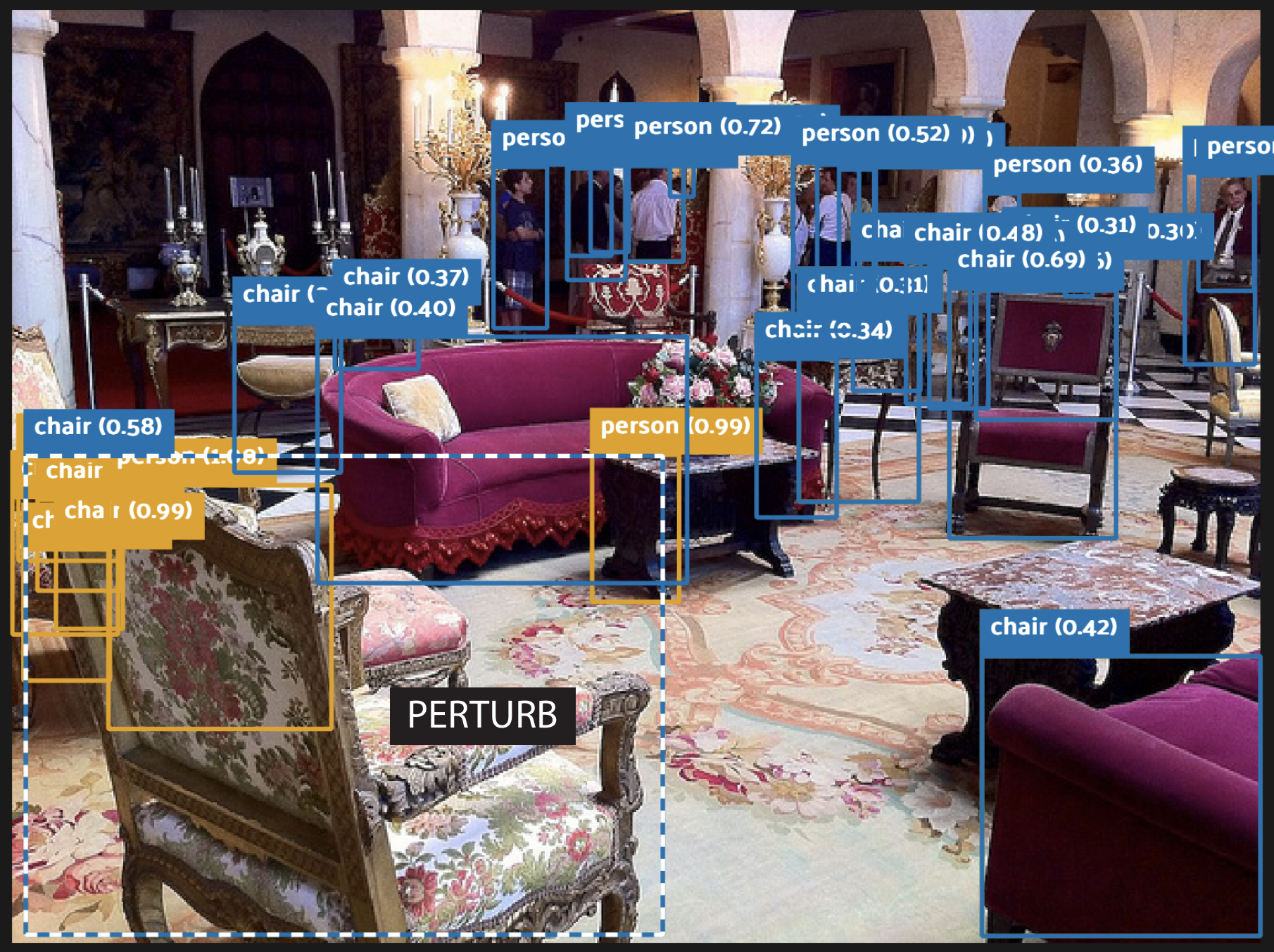} 
}

\caption{We can increase success rates by intentionally selecting target and perturb objects: An untargeted attack perturbs a large chair (dotted blue box) and causes RetinaNet to misplace the locations of people and chairs. Predictions on the original image are shown in blue and those on the adversarial image are shown in orange. For simplicity, only detections of people and chairs are shown. The corresponding perturbed image is shown in Figure \ref{fig:del_per_pert} in the appendix. }\label{fig:del_per_exp}
\end{figure}

We test all combinations. For every combination, we resample 200 COCO test images and run the 3 attacks for 200 iterations.

\textbf{Hypotheses} We tested the 3 success factors in Section \ref{sec:rand_res} and all are shown to individually increase success rates. Now we hypothesize that these success factors \textit{independently} increase success rates (i.e., success rates increase as the number of factors combined increases).

\textbf{Results} As shown in Figure \ref{fig:biased_trend_graph}, success rates increase as the number of factors used in combination increases. The attacker who includes all 3 factors obtains for the vanishing and mislabeling attacks more than 90\% success on YOLOv3 and more than 70\% success on SSD, and for the untargeted attack more than 60\% success on RetinaNet, Faster R-CNN and Cascade R-CNN. A success example is illustrated in Figure \ref{fig:del_per_exp}. 
Imposing a 0.05 $l_{\infty}$-norm constraint slightly decreases success, as shown in Figure \ref{fig:biased_trend_graph_normed} in the appendix. Since the trends remain the same, we will only conduct hypothesis testing based on the results without norm constraint.  
Hypothesis testing is similar to the procedure in the randomized experiment (Sections \ref{sec:rand_hp} and \ref{sec:rand_res}). A logistic regression model shows that success rates significantly increase as more factors are combined to select target and perturb objects for all models and attacks. Statistics are given in Table \ref{tab:num_cri_table} in the appendix.

\subsection{Perturbing Arbitrary Regions}\label{sec:del_arb}

When a perturbed object could not be selected easily, the attacker can also perturb an arbitrary region in the image to obfuscate intent. 

\begin{figure}[htb]
{\centering \includegraphics[width=0.99\columnwidth]{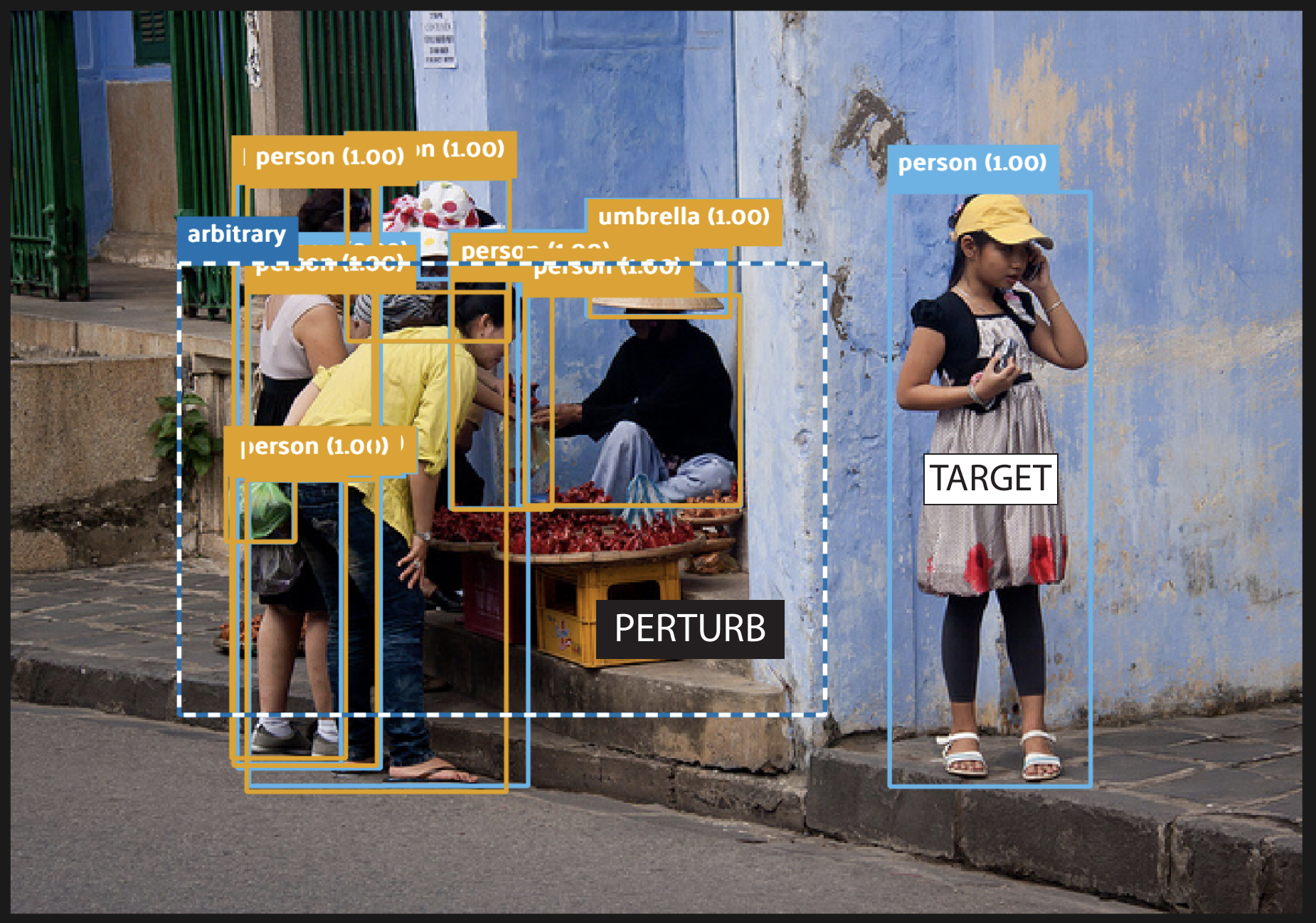}
} 

\caption{We can implement an intent obfuscating attack via perturbing an arbitrary region rather than an actual object: A mislabeling attack on Cascade R-CNN perturbs a non-overlapping arbitrary region (dotted blue) and causes the targeted person to vanish. Predictions on the original image are shown in blue and those on the adversarial image are shown in orange. The corresponding perturbed image is shown in Figure \ref{fig:del_arb_pert} in the appendix. }\label{fig:del_arb_exp}
\end{figure}

\textbf{Setup} We adopt the setup in the randomized attack (Section \ref{sec:rand_set}). However, rather than randomly selecting target and perturb objects, we randomly select a target object and then enclose a non-overlapping square perturb region beside it. We vary the length of the square perturb region to be 10, 30, 50, and 70\% of the image width or height, and vary the distance between the target and perturb bounding boxes to be 1, 5, 10, and 20\% of the image width or height. More details are given in Figure \ref{fig:del_arb_pic} in the appendix. We test all combinations. For every combination, we resample 200 COCO test images and run the 3 attacks for 200 iterations.

\textbf{Hypotheses} Actively manipulating only the perturb sizes and target-perturb distances makes the deliberate attack more controlled than the randomized attack. Hence, although we are proposing similar hypotheses to those in the randomized attack (Hypotheses \ref{hp:perturb_bbox} and \ref{hp:object_dist}), we can more strongly claim that larger perturb sizes or shorter distances \textit{cause} 
success rates to increase. 

\textbf{Results} Success rates greatly increase compared to the randomized attack (Figure \ref{fig:arbitrary_trend_graph}): when perturb lengths are more than 50\% of the image length and perturb-target distances are less than 5\% of the image length, the attacker obtains for the vanishing attack nearly 100\% success rates on YOLOv3 and SSD, and for the untargeted attack more than 25\% on RetinaNet, Faster R-CNN and Cascade R-CNN. Imposing a 0.05 $l_{\infty}$-norm constraint slightly decreases success, as shown in Figure \ref{fig:arbitrary_trend_graph_normed} in the appendix, but it is still greater than the randomized attack. A success example is illustrated in Figure \ref{fig:del_arb_exp}. Since the trends remain the same, we will only conduct hypothesis testing based on the results without norm constraint, similar to the previous two experiments.

Hypothesis testing is similar to the procedure in the randomized experiment (Sections \ref{sec:rand_hp} and \ref{sec:rand_res}): A logistic regression model using both terms as predictors show that longer perturb lengths and shorter perturb-target distances cause success rates to increase significantly for all model and attack combinations. Statistics are given in Table \ref{tab:arbitrary_trend_table} in the appendix.

\section{Discussion and Conclusion}

\textbf{Perturbing objects versus non-objects:} For intent obfuscating attacks, perturbing actual objects is intuitively more misleading than perturbing non-objects, and there is no a priori reason to believe that either will change success rates. Should the attacker then always perturb objects rather than non-objects? Surprisingly no: hypothesis testing showed that perturbing an object (in the randomized attack) rather than a non-object (in the deliberate attack) significantly \textit{decreases} success rates for most model and attack combinations, after controlling for perturb sizes and perturb-target distances, as shown in Table~\ref{tab:rand_arb_compare_table} in the appendix. Interestingly, while intent obfuscation is possible, it is more difficult to achieve than a mere contextual attack. 

\textbf{Limitations:} We have shown that intent obfuscating attacks are feasible for the 5 prominent object detectors and analyzed 10 success factors. Although we did not conduct experiments in which the attacker has no access to the victim detector, we believe that the breadth and depth of the paper will illuminate the success characteristics of intent obfuscating attacks in both settings. Interested readers can turn to \citet{Cai2021-um} for black-box contextual attacks and \citet{Lee2019-rz} for physical contextual attacks.

\section{Broader Impact} 
We have demonstrated that a malicious actor can use an intent obfuscating attack to disrupt AI systems while maintaining plausible deniability. An intent obfuscating attack goes beyond a mere contextual attack. By carefully selecting non-overlapping target and perturb regions, the malicious actor can deceive a human detective into believing their actions were innocuous.

A key defense against the attack is to use 2-stage detectors like Faster R-CNN and Cascade R-CNN. These models are shown to be more robust than 1-stage detectors like YOLOv3 and SSD against all three attacks. 
Indeed, whether to use 1-stage or 2-stage detectors is not only a matter of speed or accuracy; machine learning engineers also have to consider whether the increased resilience against intent obfuscating attacks makes 2-stage detectors more suitable, particularly in security-critical applications.

Besides the technical recommendation, we would like to raise an important legal concern: there is hardly any legal protection against intent obfuscating attacks. Established cybersecurity laws (like the United States CFAA) do not address adversarial machine learning explicitly \cite{Kumar2018-vm, Kumar2020-qi}. Intent obfuscation attacks only compound the problem, since proving malicious intent is required for criminal prosecution \citep{Wex_Definitions_Team2024-ow}. To conclude, we believe that establishing the feasibility of intent obfuscating attacks will galvanize the machine learning community to develop more robust technical and legal solutions.

\section{Code and Data}

The code is available on the github repository \url{https://github.com/zhaobin-li/intent-obfusc}. The included README.md contains instructions to reproduce graphs and tables, download datasets and images, visualize attacked datasets, and replicate experiments. The datasets and perturbed images in both experiments are stored on a Google Cloud Storage bucket \url{https://console.cloud.google.com/storage/browser/intent-obfusc} (you will still need to sign in with a google account simply to access the public bucket). 

\section*{Acknowledgements}
We thank Scott Cheng-Hsin Yang and Wei-Ting Chiu for editing the paper. This work was supported in part by a grant from the DARPA RED program (20-430 Rev00-NJ-112) to PS.

\bibliography{paperpile.bib,packages.bib}

\appendix
\onecolumn

\section{Table Headers}\label{app:tab_hdr}
We generate the graphs and tables in the sections below using R \citep{R-base}. The upper table headers are generated using R knitr \citep{R-knitr} and kableExtra \citep{R-kableExtra}: We run one regression per group (model and/or attack combination). The terms with a blank row and 0.000 estimate are reference variables in the regression model, e.g.\ YOLOv3 in Table \ref{tab:model_stage_table}. The lower regression headers are generated using R broom \citep{R-broom} and broom.helpers \citep{R-broom.helpers}. To adapt the broom documentation at \url{https://broom.tidymodels.org/reference/tidy.lm.html#value}:

\begin{description}
    \item[term] The name of the regression term.
    \item[sig] Terms which are significant ($p < .05$) are denoted by ``*''.
    \item[estimate] The estimated value of the regression term.
    \item[std.error] The standard error of the regression term.
    \item[statistic]The value of a Wald z-statistic to use in a hypothesis that the regression term is non-zero.
    \item[p.value] The two-sided p-value associated with the observed statistic.
    \item[conf.low] Lower bound on the 95\% confidence interval for the estimate.
    \item[conf.high] Upper bound on the 95\% confidence interval for the estimate.
\end{description}

\section{Randomized Attack}\label{app:rand_att}

\subsection{Setup}\label{app:rand_set}
Since we are using a shared computing resource on an internal network, we split the attack into 20 repetitions and attacked 200 images per repetition. The images are randomly sampled without replacement within repetitions, but may repeat across repetitions. Every repetition takes approximately 60 minutes on a 32GB NVIDIA Tesla V100 GPU. 2400 repetitions (5 models * 3 attacks * 4 iterations * 20 repetitions * 2 norms) take 100 V100 GPU days. More complex models (e.g. Cascade R-CNN) require more attack time than less complex models (e.g. YOLOv3).

Across model, attack and iteration combinations, we sample the same images and select the same target and perturb objects per image to more accurately compare the success rates between combinations. In addition, the MMdetection models backpropagate only in training mode. Hence, we set the model to training mode in the TOG attack to backpropagate the gradients. Since the model evaluates the adversarial images in testing mode, we reset the model after every iteration to prevent updates to its weights or running statistics, to ensure the gradients are directed towards the model in testing mode. Also, we do not use data augmentation in the TOG attack, since the adversarial images are not augmented during evaluation.

\subsection{Results}\label{app:rand_tab}
\begingroup\fontsize{9}{11}\selectfont

\begin{longtable}[t]{lllrrrrrr}
\caption{\label{tab:model_stage_table}We run a logistic model regressing success against detection models, split by attack, in the randomized attack experiment. Both vanishing and mislabeling attacks obtain higher success on 1-stage (YOLOv3, SSD) than 2-stage (Faster R-CNN, Cascade R-CNN) detectors. However, the 1-stage RetinaNet is as resilient as 2-stage detectors. Table headers are explained in Appendix \ref{app:tab_hdr}.}\\
\toprule
\multicolumn{1}{c}{Group} & \multicolumn{8}{c}{Regression} \\
\cmidrule(l{3pt}r{3pt}){1-1} \cmidrule(l{3pt}r{3pt}){2-9}
Attack & term & sig & estimate & std.error & statistic & p.value & conf.low & conf.high\\
\midrule
 & YOLOv3 &  & 0.000 &  &  &  &  & \\
\cmidrule{2-9}\nopagebreak
 & SSD &  & -0.029 & 0.048 & -0.597 & 0.550 & -0.122 & 0.065\\
\cmidrule{2-9}\nopagebreak
 & RetinaNet & * & -1.685 & 0.067 & -25.317 & 0.000 & -1.817 & -1.556\\
\cmidrule{2-9}\nopagebreak
 & Faster R-CNN & * & -2.352 & 0.084 & -28.021 & 0.000 & -2.519 & -2.190\\
\cmidrule{2-9}\nopagebreak
\multirow{-5}{*}{\raggedright\arraybackslash Vanishing} & Cascade R-CNN & * & -1.929 & 0.072 & -26.776 & 0.000 & -2.072 & -1.790\\
\cmidrule{1-9}\pagebreak[0]
 & YOLOv3 &  & 0.000 &  &  &  &  & \\
\cmidrule{2-9}\nopagebreak
 & SSD & * & 0.361 & 0.058 & 6.239 & 0.000 & 0.248 & 0.475\\
\cmidrule{2-9}\nopagebreak
 & RetinaNet & * & -2.052 & 0.112 & -18.248 & 0.000 & -2.278 & -1.837\\
\cmidrule{2-9}\nopagebreak
 & Faster R-CNN & * & -2.555 & 0.139 & -18.371 & 0.000 & -2.838 & -2.292\\
\cmidrule{2-9}\nopagebreak
\multirow{-5}{*}{\raggedright\arraybackslash Mislabeling} & Cascade R-CNN & * & -1.706 & 0.098 & -17.372 & 0.000 & -1.902 & -1.517\\
\cmidrule{1-9}\pagebreak[0]
 & YOLOv3 &  & 0.000 &  &  &  &  & \\
\cmidrule{2-9}\nopagebreak
 & SSD & * & 1.123 & 0.068 & 16.407 & 0.000 & 0.990 & 1.258\\
\cmidrule{2-9}\nopagebreak
 & RetinaNet &  & 0.084 & 0.079 & 1.066 & 0.286 & -0.071 & 0.239\\
\cmidrule{2-9}\nopagebreak
 & Faster R-CNN &  & 0.099 & 0.079 & 1.259 & 0.208 & -0.055 & 0.254\\
\cmidrule{2-9}\nopagebreak
\multirow{-5}{*}{\raggedright\arraybackslash Untargeted} & Cascade R-CNN & * & -0.304 & 0.086 & -3.531 & 0.000 & -0.474 & -0.136\\
\bottomrule
\end{longtable}
\endgroup{}

\begingroup\fontsize{9}{11}\selectfont

\begin{longtable}[t]{lllrrrrrr}
\caption{\label{tab:target_untarget_vanish_mislabel_table}We run a logistic model regressing success against attacks, split by detection models in the randomized attack experiment. Targeted attacks obtain higher success than untargeted attacks on YOLOv3 and SSD. Within targeted attacks, vanishing attacks obtain higher success than mislabeling attacks on all models. Table headers are explained in Appendix \ref{app:tab_hdr}.}\\
\toprule
\multicolumn{1}{c}{Group} & \multicolumn{8}{c}{Regression} \\
\cmidrule(l{3pt}r{3pt}){1-1} \cmidrule(l{3pt}r{3pt}){2-9}
Model & term & sig & estimate & std.error & statistic & p.value & conf.low & conf.high\\
\midrule
 & Vanishing &  & 0.000 &  &  &  &  & \\
\cmidrule{2-9}\nopagebreak
 & Mislabeling & * & -0.943 & 0.055 & -17.212 & 0.000 & -1.051 & -0.836\\
\cmidrule{2-9}\nopagebreak
\multirow{-3}{*}{\raggedright\arraybackslash YOLOv3} & Untargeted & * & -1.662 & 0.066 & -25.151 & 0.000 & -1.793 & -1.534\\
\cmidrule{1-9}\pagebreak[0]
 & Vanishing &  & 0.000 &  &  &  &  & \\
\cmidrule{2-9}\nopagebreak
 & Mislabeling & * & -0.553 & 0.051 & -10.779 & 0.000 & -0.654 & -0.453\\
\cmidrule{2-9}\nopagebreak
\multirow{-3}{*}{\raggedright\arraybackslash SSD} & Untargeted & * & -0.511 & 0.051 & -10.017 & 0.000 & -0.611 & -0.411\\
\cmidrule{1-9}\pagebreak[0]
 & Vanishing &  & 0.000 &  &  &  &  & \\
\cmidrule{2-9}\nopagebreak
 & Mislabeling & * & -1.311 & 0.119 & -11.047 & 0.000 & -1.548 & -1.082\\
\cmidrule{2-9}\nopagebreak
\multirow{-3}{*}{\raggedright\arraybackslash RetinaNet} & Untargeted &  & 0.107 & 0.079 & 1.348 & 0.178 & -0.048 & 0.263\\
\cmidrule{1-9}\pagebreak[0]
 & Vanishing &  & 0.000 &  &  &  &  & \\
\cmidrule{2-9}\nopagebreak
 & Mislabeling & * & -1.146 & 0.153 & -7.493 & 0.000 & -1.454 & -0.853\\
\cmidrule{2-9}\nopagebreak
\multirow{-3}{*}{\raggedright\arraybackslash Faster R-CNN} & Untargeted & * & 0.789 & 0.094 & 8.370 & 0.000 & 0.606 & 0.976\\
\cmidrule{1-9}\pagebreak[0]
 & Vanishing &  & 0.000 &  &  &  &  & \\
\cmidrule{2-9}\nopagebreak
 & Mislabeling & * & -0.720 & 0.109 & -6.619 & 0.000 & -0.936 & -0.509\\
\cmidrule{2-9}\nopagebreak
\multirow{-3}{*}{\raggedright\arraybackslash Cascade R-CNN} & Untargeted &  & -0.037 & 0.091 & -0.409 & 0.683 & -0.215 & 0.141\\
\bottomrule
\end{longtable}
\endgroup{}

\begingroup\fontsize{9}{11}\selectfont

\begin{longtable}[t]{llllrrrrrr}
\caption{\label{tab:num_iteration_table}We run a logistic model regressing success against log(attack iterations) in the randomized attack experiment. Success rates increase with attack iterations for all models and attacks. Table headers are explained in Appendix \ref{app:tab_hdr}.}\\
\toprule
\multicolumn{2}{c}{Group} & \multicolumn{8}{c}{Regression} \\
\cmidrule(l{3pt}r{3pt}){1-2} \cmidrule(l{3pt}r{3pt}){3-10}
 & Attack & term & sig & estimate & std.error & statistic & p.value & conf.low & conf.high\\
\midrule
\addlinespace[0.3em]
\multicolumn{10}{l}{\textbf{YOLOv3}}\\
\hspace{1em} & Vanishing & log(iterations) & * & 0.476 & 0.019 & 25.267 & 0 & 0.439 & 0.513\\
\cmidrule{2-10}\nopagebreak
\hspace{1em} & Mislabeling & log(iterations) & * & 0.622 & 0.030 & 20.761 & 0 & 0.564 & 0.681\\
\cmidrule{2-10}\nopagebreak
\hspace{1em} & Untargeted & log(iterations) & * & 0.192 & 0.028 & 6.776 & 0 & 0.137 & 0.247\\
\cmidrule{1-10}\pagebreak[0]
\addlinespace[0.3em]
\multicolumn{10}{l}{\textbf{SSD}}\\
\hspace{1em} & Vanishing & log(iterations) & * & 0.566 & 0.020 & 28.456 & 0 & 0.527 & 0.605\\
\cmidrule{2-10}\nopagebreak
\hspace{1em} & Mislabeling & log(iterations) & * & 0.621 & 0.025 & 24.466 & 0 & 0.572 & 0.672\\
\cmidrule{2-10}\nopagebreak
\hspace{1em} & Untargeted & log(iterations) & * & 0.256 & 0.019 & 13.449 & 0 & 0.219 & 0.294\\
\cmidrule{1-10}\pagebreak[0]
\addlinespace[0.3em]
\multicolumn{10}{l}{\textbf{RetinaNet}}\\
\hspace{1em} & Vanishing & log(iterations) & * & 0.467 & 0.037 & 12.620 & 0 & 0.396 & 0.541\\
\cmidrule{2-10}\nopagebreak
\hspace{1em} & Mislabeling & log(iterations) & * & 0.635 & 0.076 & 8.331 & 0 & 0.490 & 0.789\\
\cmidrule{2-10}\nopagebreak
\hspace{1em} & Untargeted & log(iterations) & * & 0.225 & 0.029 & 7.802 & 0 & 0.169 & 0.282\\
\cmidrule{1-10}\pagebreak[0]
\addlinespace[0.3em]
\multicolumn{10}{l}{\textbf{Faster R-CNN}}\\
\hspace{1em} & Vanishing & log(iterations) & * & 0.397 & 0.049 & 8.160 & 0 & 0.303 & 0.494\\
\cmidrule{2-10}\nopagebreak
\hspace{1em} & Mislabeling & log(iterations) & * & 0.534 & 0.093 & 5.762 & 0 & 0.358 & 0.722\\
\cmidrule{2-10}\nopagebreak
\hspace{1em} & Untargeted & log(iterations) & * & 0.367 & 0.034 & 10.897 & 0 & 0.302 & 0.434\\
\cmidrule{1-10}\pagebreak[0]
\addlinespace[0.3em]
\multicolumn{10}{l}{\textbf{Cascade R-CNN}}\\
\hspace{1em} & Vanishing & log(iterations) & * & 0.502 & 0.043 & 11.736 & 0 & 0.419 & 0.587\\
\cmidrule{2-10}\nopagebreak
\hspace{1em} & Mislabeling & log(iterations) & * & 0.753 & 0.073 & 10.276 & 0 & 0.613 & 0.901\\
\cmidrule{2-10}\nopagebreak
\hspace{1em} & Untargeted & log(iterations) & * & 0.325 & 0.038 & 8.477 & 0 & 0.251 & 0.401\\
\bottomrule
\end{longtable}
\endgroup{}

\section{Analyze individual cases}\label{analyze-individual-cases}

\begin{figure}[tb]

{\centering \includegraphics{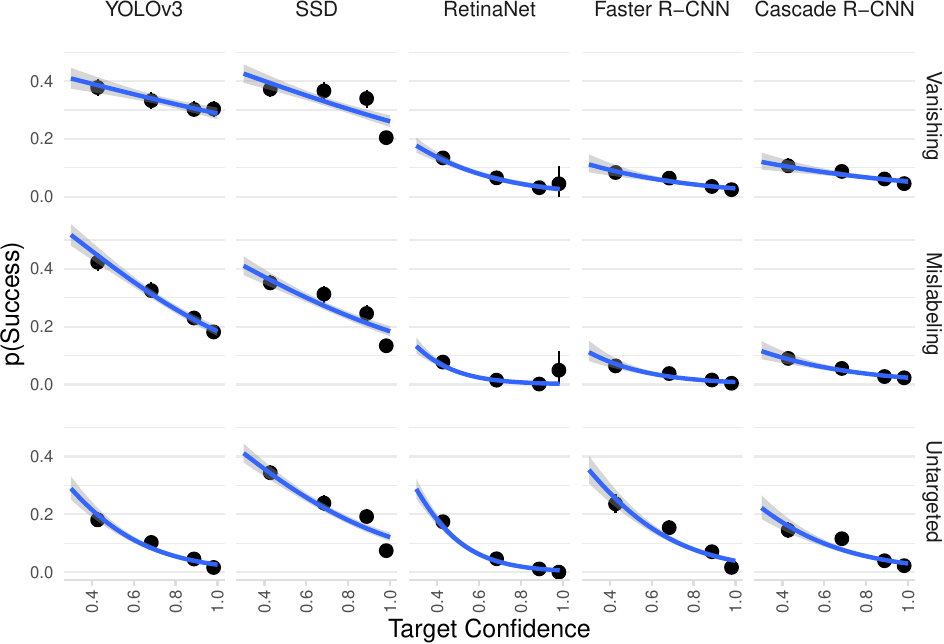} 

}

\caption{Lower target confidence significantly increases success rates for all models and attacks:  The binned summaries and regression trendlines graph success proportion against target confidence in the randomized attack experiment. Bins are split into quantiles. Errors are 95\% confidence intervals}\label{fig:target_conf_graph}
\end{figure}

\begingroup\fontsize{9}{11}\selectfont

\begin{longtable}[t]{llllrrrrrr}
\caption{\label{tab:target_conf_table}We run a logistic model regressing success against target confidence in the randomized attack experiment. Lower target confidence significantly increases success rates for all models and attacks. Table headers are explained in Appendix \ref{app:tab_hdr}.}\\
\toprule
\multicolumn{2}{c}{Group} & \multicolumn{8}{c}{Regression} \\
\cmidrule(l{3pt}r{3pt}){1-2} \cmidrule(l{3pt}r{3pt}){3-10}
 & Attack & term & sig & estimate & std.error & statistic & p.value & conf.low & conf.high\\
\midrule
\addlinespace[0.3em]
\multicolumn{10}{l}{\textbf{YOLOv3}}\\
\hspace{1em} & Vanishing & confidence & * & -0.773 & 0.153 & -5.059 & 0 & -1.072 & -0.473\\
\cmidrule{2-10}\nopagebreak
\hspace{1em} & Mislabeling & confidence & * & -2.230 & 0.160 & -13.915 & 0 & -2.545 & -1.917\\
\cmidrule{2-10}\nopagebreak
\hspace{1em} & Untargeted & confidence & * & -3.910 & 0.268 & -14.579 & 0 & -4.442 & -3.390\\
\cmidrule{1-10}\pagebreak[0]
\addlinespace[0.3em]
\multicolumn{10}{l}{\textbf{SSD}}\\
\hspace{1em} & Vanishing & confidence & * & -1.063 & 0.142 & -7.505 & 0 & -1.341 & -0.786\\
\cmidrule{2-10}\nopagebreak
\hspace{1em} & Mislabeling & confidence & * & -1.616 & 0.151 & -10.714 & 0 & -1.913 & -1.321\\
\cmidrule{2-10}\nopagebreak
\hspace{1em} & Untargeted & confidence & * & -2.326 & 0.164 & -14.203 & 0 & -2.649 & -2.007\\
\cmidrule{1-10}\pagebreak[0]
\addlinespace[0.3em]
\multicolumn{10}{l}{\textbf{RetinaNet}}\\
\hspace{1em} & Vanishing & confidence & * & -3.057 & 0.321 & -9.535 & 0 & -3.695 & -2.437\\
\cmidrule{2-10}\nopagebreak
\hspace{1em} & Mislabeling & confidence & * & -6.133 & 0.616 & -9.952 & 0 & -7.389 & -4.969\\
\cmidrule{2-10}\nopagebreak
\hspace{1em} & Untargeted & confidence & * & -6.050 & 0.400 & -15.130 & 0 & -6.853 & -5.284\\
\cmidrule{1-10}\pagebreak[0]
\addlinespace[0.3em]
\multicolumn{10}{l}{\textbf{Faster R-CNN}}\\
\hspace{1em} & Vanishing & confidence & * & -2.079 & 0.326 & -6.383 & 0 & -2.714 & -1.436\\
\cmidrule{2-10}\nopagebreak
\hspace{1em} & Mislabeling & confidence & * & -3.903 & 0.449 & -8.702 & 0 & -4.795 & -3.032\\
\cmidrule{2-10}\nopagebreak
\hspace{1em} & Untargeted & confidence & * & -3.719 & 0.239 & -15.564 & 0 & -4.190 & -3.253\\
\cmidrule{1-10}\pagebreak[0]
\addlinespace[0.3em]
\multicolumn{10}{l}{\textbf{Cascade R-CNN}}\\
\hspace{1em} & Vanishing & confidence & * & -1.298 & 0.275 & -4.727 & 0 & -1.831 & -0.754\\
\cmidrule{2-10}\nopagebreak
\hspace{1em} & Mislabeling & confidence & * & -2.428 & 0.332 & -7.317 & 0 & -3.077 & -1.775\\
\cmidrule{2-10}\nopagebreak
\hspace{1em} & Untargeted & confidence & * & -3.183 & 0.271 & -11.740 & 0 & -3.716 & -2.653\\
\bottomrule
\end{longtable}
\endgroup{}

\begin{figure}[tb]

{\centering \includegraphics{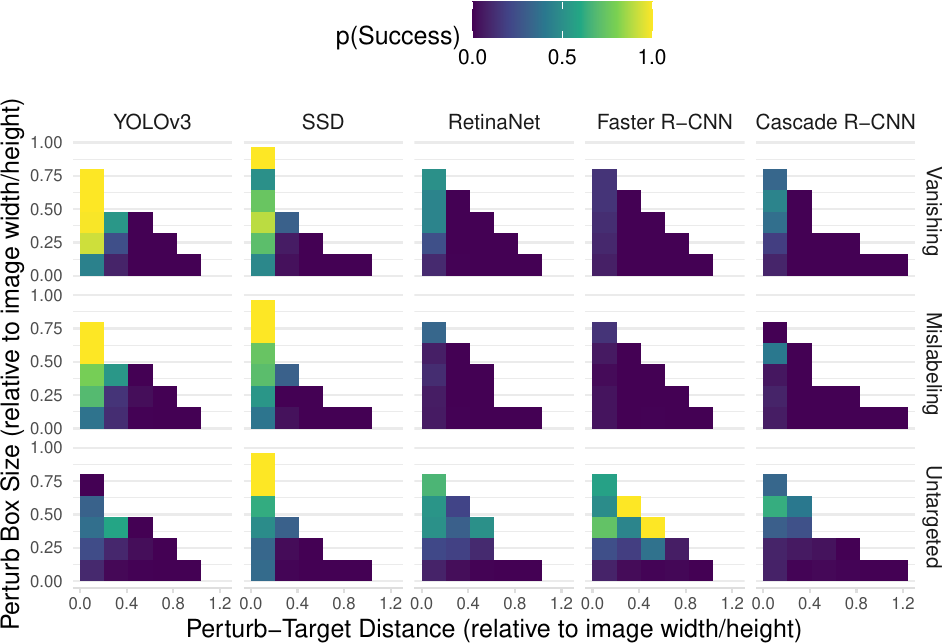} 

}

\caption{Larger perturb objects significantly increase success rates for all models and attacks, except for mislabeling attack on Faster R-CNN, after controlling for perturb-target distances. Shorter perturb-target distances significantly increase success rates for all models and attacks, after controlling for perturb object sizes:  The binned summaries graph success proportion against perturb-target distance (relative to image width/height) and perturb box size (relative to image width/height) in the randomized attack experiment.}\label{fig:perturb_bbox_and_object_dist_graph}
\end{figure}

\begingroup\fontsize{9}{11}\selectfont

\begin{longtable}[t]{llllrrrrrr}
\caption{\label{tab:perturb_bbox_and_object_dist_table}We run a logistic model regressing success against perturb-target distance (relative to image width/height) and perturb box size (relative to image width/height) in the randomized attack experiment. Larger perturb objects significantly increase success rates for all models and attacks, except for mislabeling attack on Faster R-CNN, after controlling for perturb-target distances. Shorter perturb-target distances significantly increase success rates for all models and attacks, after controlling for perturb object sizes. Table headers are explained in Appendix \ref{app:tab_hdr}.}\\
\toprule
\multicolumn{2}{c}{Group} & \multicolumn{8}{c}{Regression} \\
\cmidrule(l{3pt}r{3pt}){1-2} \cmidrule(l{3pt}r{3pt}){3-10}
 & Attack & term & sig & estimate & std.error & statistic & p.value & conf.low & conf.high\\
\midrule
\addlinespace[0.3em]
\multicolumn{10}{l}{\textbf{YOLOv3}}\\
\hspace{1em} & Vanishing & distance & * & -9.672 & 0.656 & -14.738 & 0.000 & -10.986 & -8.413\\
\cmidrule{3-10}\nopagebreak
\hspace{1em} &  & size & * & 32.877 & 2.200 & 14.945 & 0.000 & 28.697 & 37.320\\
\cmidrule{3-10}\nopagebreak
\hspace{1em} &  & distance * size & * & -96.578 & 10.405 & -9.282 & 0.000 & -117.509 & -76.730\\
\cmidrule{2-10}\nopagebreak
\hspace{1em} & Mislabeling & distance & * & -8.322 & 0.516 & -16.121 & 0.000 & -9.355 & -7.331\\
\cmidrule{3-10}\nopagebreak
\hspace{1em} &  & size & * & 8.229 & 0.837 & 9.833 & 0.000 & 6.635 & 9.917\\
\cmidrule{3-10}\nopagebreak
\hspace{1em} &  & distance * size & * & -9.864 & 4.876 & -2.023 & 0.043 & -19.658 & -0.531\\
\cmidrule{2-10}\nopagebreak
\hspace{1em} & Untargeted & distance & * & -13.317 & 1.151 & -11.566 & 0.000 & -15.649 & -11.136\\
\cmidrule{3-10}\nopagebreak
\hspace{1em} &  & size & * & 1.638 & 0.647 & 2.532 & 0.011 & 0.369 & 2.909\\
\cmidrule{3-10}\nopagebreak
\hspace{1em} &  & distance * size & * & 31.584 & 5.862 & 5.388 & 0.000 & 20.028 & 43.048\\
\cmidrule{1-10}\pagebreak[0]
\addlinespace[0.3em]
\multicolumn{10}{l}{\textbf{SSD}}\\
\hspace{1em} & Vanishing & distance & * & -14.374 & 0.758 & -18.971 & 0.000 & -15.892 & -12.921\\
\cmidrule{3-10}\nopagebreak
\hspace{1em} &  & size & * & 9.330 & 0.959 & 9.729 & 0.000 & 7.508 & 11.267\\
\cmidrule{3-10}\nopagebreak
\hspace{1em} &  & distance * size &  & -7.647 & 5.626 & -1.359 & 0.174 & -18.998 & 3.079\\
\cmidrule{2-10}\nopagebreak
\hspace{1em} & Mislabeling & distance & * & -12.008 & 0.729 & -16.468 & 0.000 & -13.473 & -10.614\\
\cmidrule{3-10}\nopagebreak
\hspace{1em} &  & size & * & 7.727 & 0.806 & 9.591 & 0.000 & 6.198 & 9.357\\
\cmidrule{3-10}\nopagebreak
\hspace{1em} &  & distance * size & * & -13.614 & 5.556 & -2.451 & 0.014 & -24.820 & -3.030\\
\cmidrule{2-10}\nopagebreak
\hspace{1em} & Untargeted & distance & * & -14.125 & 0.811 & -17.425 & 0.000 & -15.757 & -12.579\\
\cmidrule{3-10}\nopagebreak
\hspace{1em} &  & size & * & 2.298 & 0.528 & 4.353 & 0.000 & 1.289 & 3.361\\
\cmidrule{3-10}\nopagebreak
\hspace{1em} &  & distance * size & * & 11.937 & 4.573 & 2.611 & 0.009 & 2.779 & 20.724\\
\cmidrule{1-10}\pagebreak[0]
\addlinespace[0.3em]
\multicolumn{10}{l}{\textbf{RetinaNet}}\\
\hspace{1em} & Vanishing & distance & * & -38.670 & 2.842 & -13.608 & 0.000 & -44.429 & -33.288\\
\cmidrule{3-10}\nopagebreak
\hspace{1em} &  & size & * & 1.917 & 0.675 & 2.840 & 0.005 & 0.647 & 3.291\\
\cmidrule{3-10}\nopagebreak
\hspace{1em} &  & distance * size & * & 53.194 & 10.742 & 4.952 & 0.000 & 31.190 & 73.157\\
\cmidrule{2-10}\nopagebreak
\hspace{1em} & Mislabeling & distance & * & -48.140 & 5.186 & -9.283 & 0.000 & -58.781 & -38.448\\
\cmidrule{3-10}\nopagebreak
\hspace{1em} &  & size & * & 2.270 & 1.151 & 1.972 & 0.049 & 0.074 & 4.594\\
\cmidrule{3-10}\nopagebreak
\hspace{1em} &  & distance * size &  & 7.234 & 25.556 & 0.283 & 0.777 & -46.376 & 53.609\\
\cmidrule{2-10}\nopagebreak
\hspace{1em} & Untargeted & distance & * & -13.171 & 1.189 & -11.082 & 0.000 & -15.598 & -10.938\\
\cmidrule{3-10}\nopagebreak
\hspace{1em} &  & size & * & 2.541 & 0.519 & 4.892 & 0.000 & 1.526 & 3.565\\
\cmidrule{3-10}\nopagebreak
\hspace{1em} &  & distance * size & * & 36.039 & 4.724 & 7.629 & 0.000 & 27.007 & 45.549\\
\cmidrule{1-10}\pagebreak[0]
\addlinespace[0.3em]
\multicolumn{10}{l}{\textbf{Faster R-CNN}}\\
\hspace{1em} & Vanishing & distance & * & -31.462 & 3.270 & -9.622 & 0.000 & -38.181 & -25.358\\
\cmidrule{3-10}\nopagebreak
\hspace{1em} &  & size & * & 3.758 & 1.086 & 3.462 & 0.001 & 1.675 & 5.942\\
\cmidrule{3-10}\nopagebreak
\hspace{1em} &  & distance * size &  & -35.320 & 23.347 & -1.513 & 0.130 & -84.636 & 7.187\\
\cmidrule{2-10}\nopagebreak
\hspace{1em} & Mislabeling & distance & * & -24.289 & 3.513 & -6.914 & 0.000 & -31.624 & -17.853\\
\cmidrule{3-10}\nopagebreak
\hspace{1em} &  & size &  & 1.648 & 1.414 & 1.166 & 0.244 & -1.207 & 4.385\\
\cmidrule{3-10}\nopagebreak
\hspace{1em} &  & distance * size &  & -37.467 & 32.660 & -1.147 & 0.251 & -108.916 & 19.888\\
\cmidrule{2-10}\nopagebreak
\hspace{1em} & Untargeted & distance & * & -14.429 & 1.244 & -11.603 & 0.000 & -16.949 & -12.074\\
\cmidrule{3-10}\nopagebreak
\hspace{1em} &  & size & * & 2.184 & 0.650 & 3.360 & 0.001 & 0.913 & 3.465\\
\cmidrule{3-10}\nopagebreak
\hspace{1em} &  & distance * size & * & 58.694 & 5.959 & 9.849 & 0.000 & 47.273 & 70.648\\
\cmidrule{1-10}\pagebreak[0]
\addlinespace[0.3em]
\multicolumn{10}{l}{\textbf{Cascade R-CNN}}\\
\hspace{1em} & Vanishing & distance & * & -27.740 & 2.837 & -9.778 & 0.000 & -33.578 & -22.453\\
\cmidrule{3-10}\nopagebreak
\hspace{1em} &  & size & * & 7.189 & 0.906 & 7.936 & 0.000 & 5.488 & 9.045\\
\cmidrule{3-10}\nopagebreak
\hspace{1em} &  & distance * size & * & -77.368 & 22.567 & -3.428 & 0.001 & -125.142 & -36.519\\
\cmidrule{2-10}\nopagebreak
\hspace{1em} & Mislabeling & distance & * & -28.681 & 3.361 & -8.533 & 0.000 & -35.680 & -22.493\\
\cmidrule{3-10}\nopagebreak
\hspace{1em} &  & size & * & 2.584 & 0.763 & 3.388 & 0.001 & 1.094 & 4.093\\
\cmidrule{3-10}\nopagebreak
\hspace{1em} &  & distance * size & * & -69.647 & 31.193 & -2.233 & 0.026 & -136.025 & -13.985\\
\cmidrule{2-10}\nopagebreak
\hspace{1em} & Untargeted & distance & * & -13.415 & 1.297 & -10.340 & 0.000 & -16.058 & -10.972\\
\cmidrule{3-10}\nopagebreak
\hspace{1em} &  & size & * & 2.594 & 0.561 & 4.621 & 0.000 & 1.492 & 3.697\\
\cmidrule{3-10}\nopagebreak
\hspace{1em} &  & distance * size & * & 25.276 & 4.976 & 5.079 & 0.000 & 15.453 & 35.061\\
\bottomrule
\end{longtable}
\endgroup{}

\begin{figure}[tb]

{\centering \includegraphics{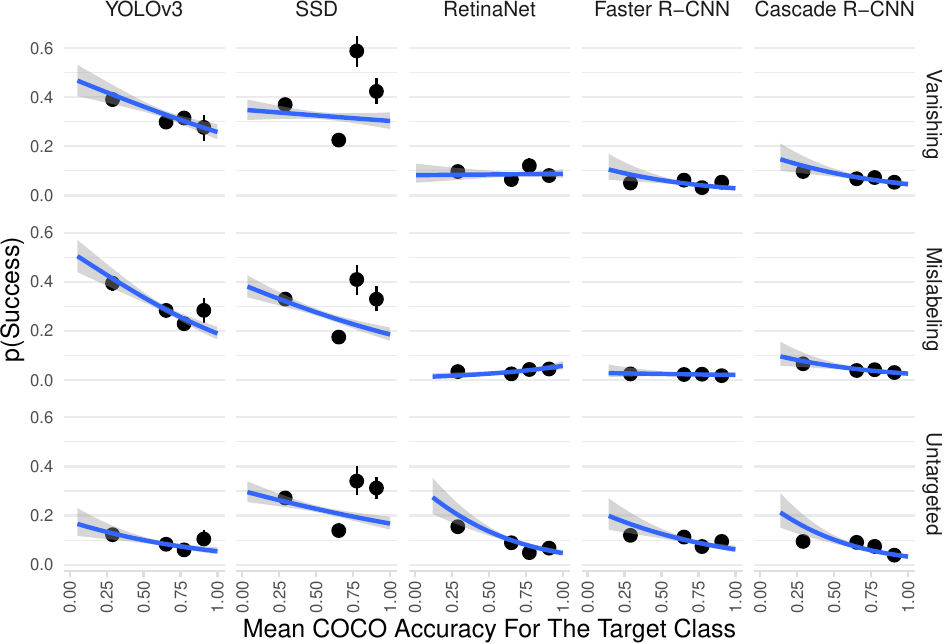} 

}

\caption{Although higher mean COCO accuracy for the target class seem to decrease success rates, the results are mixed after controlling for target class confidence (Table \ref{tab:target_success_table}):  The binned summaries and regression trendlines graph success proportion against mean COCO accuracy for the target class in the randomized attack experiment. Bins are split into quantiles. Errors are 95\% confidence intervals}\label{fig:target_success_graph}
\end{figure}

\begingroup\fontsize{9}{11}\selectfont

\begin{longtable}[t]{llllrrrrrr}
\caption{\label{tab:target_success_table}We run a logistic model regressing success against mean COCO accuracy for the target class, with target confidence as covariate, in the randomized attack experiment. The results are mixed after controlling for target class confidence and the relatively large interaction terms make interpretation challenging. Table headers are explained in Appendix \ref{app:tab_hdr}.}\\
\toprule
\multicolumn{2}{c}{Group} & \multicolumn{8}{c}{Regression} \\
\cmidrule(l{3pt}r{3pt}){1-2} \cmidrule(l{3pt}r{3pt}){3-10}
 & Attack & term & sig & estimate & std.error & statistic & p.value & conf.low & conf.high\\
\midrule
\addlinespace[0.3em]
\multicolumn{10}{l}{\textbf{YOLOv3}}\\
\hspace{1em} & Vanishing & accuracy &  & 0.726 & 0.732 & 0.992 & 0.321 & -0.707 & 2.164\\
\cmidrule{3-10}\nopagebreak
\hspace{1em} &  & confidence &  & 0.733 & 0.652 & 1.124 & 0.261 & -0.544 & 2.014\\
\cmidrule{3-10}\nopagebreak
\hspace{1em} &  & accuracy * confidence & * & -2.196 & 0.976 & -2.250 & 0.024 & -4.113 & -0.285\\
\cmidrule{2-10}\nopagebreak
\hspace{1em} & Mislabeling & accuracy &  & 1.133 & 0.743 & 1.524 & 0.128 & -0.325 & 2.591\\
\cmidrule{3-10}\nopagebreak
\hspace{1em} &  & confidence &  & 0.044 & 0.679 & 0.065 & 0.948 & -1.289 & 1.373\\
\cmidrule{3-10}\nopagebreak
\hspace{1em} &  & accuracy * confidence & * & -3.371 & 1.025 & -3.289 & 0.001 & -5.382 & -1.363\\
\cmidrule{2-10}\nopagebreak
\hspace{1em} & Untargeted & accuracy &  & 1.324 & 1.060 & 1.248 & 0.212 & -0.749 & 3.410\\
\cmidrule{3-10}\nopagebreak
\hspace{1em} &  & confidence &  & -1.696 & 1.113 & -1.525 & 0.127 & -3.895 & 0.469\\
\cmidrule{3-10}\nopagebreak
\hspace{1em} &  & accuracy * confidence & * & -3.376 & 1.697 & -1.989 & 0.047 & -6.701 & -0.047\\
\cmidrule{1-10}\pagebreak[0]
\addlinespace[0.3em]
\multicolumn{10}{l}{\textbf{SSD}}\\
\hspace{1em} & Vanishing & accuracy & * & 1.282 & 0.511 & 2.508 & 0.012 & 0.283 & 2.288\\
\cmidrule{3-10}\nopagebreak
\hspace{1em} &  & confidence &  & 0.017 & 0.426 & 0.040 & 0.968 & -0.816 & 0.854\\
\cmidrule{3-10}\nopagebreak
\hspace{1em} &  & accuracy * confidence & * & -1.907 & 0.710 & -2.684 & 0.007 & -3.304 & -0.519\\
\cmidrule{2-10}\nopagebreak
\hspace{1em} & Mislabeling & accuracy & * & 3.281 & 0.549 & 5.976 & 0.000 & 2.210 & 4.363\\
\cmidrule{3-10}\nopagebreak
\hspace{1em} &  & confidence & * & 1.871 & 0.460 & 4.067 & 0.000 & 0.972 & 2.776\\
\cmidrule{3-10}\nopagebreak
\hspace{1em} &  & accuracy * confidence & * & -6.178 & 0.795 & -7.769 & 0.000 & -7.747 & -4.629\\
\cmidrule{2-10}\nopagebreak
\hspace{1em} & Untargeted & accuracy & * & 4.517 & 0.584 & 7.738 & 0.000 & 3.381 & 5.670\\
\cmidrule{3-10}\nopagebreak
\hspace{1em} &  & confidence & * & 1.990 & 0.499 & 3.985 & 0.000 & 1.014 & 2.971\\
\cmidrule{3-10}\nopagebreak
\hspace{1em} &  & accuracy * confidence & * & -7.783 & 0.874 & -8.905 & 0.000 & -9.508 & -6.081\\
\cmidrule{1-10}\pagebreak[0]
\addlinespace[0.3em]
\multicolumn{10}{l}{\textbf{RetinaNet}}\\
\hspace{1em} & Vanishing & accuracy &  & 1.009 & 1.143 & 0.883 & 0.377 & -1.217 & 3.262\\
\cmidrule{3-10}\nopagebreak
\hspace{1em} &  & confidence & * & -3.823 & 1.744 & -2.192 & 0.028 & -7.277 & -0.442\\
\cmidrule{3-10}\nopagebreak
\hspace{1em} &  & accuracy * confidence &  & 0.571 & 2.246 & 0.254 & 0.799 & -3.819 & 4.984\\
\cmidrule{2-10}\nopagebreak
\hspace{1em} & Mislabeling & accuracy &  & 2.565 & 2.044 & 1.255 & 0.209 & -1.385 & 6.612\\
\cmidrule{3-10}\nopagebreak
\hspace{1em} &  & confidence & * & -8.994 & 3.794 & -2.371 & 0.018 & -16.549 & -1.716\\
\cmidrule{3-10}\nopagebreak
\hspace{1em} &  & accuracy * confidence &  & 2.506 & 4.691 & 0.534 & 0.593 & -6.650 & 11.687\\
\cmidrule{2-10}\nopagebreak
\hspace{1em} & Untargeted & accuracy & * & 2.471 & 1.206 & 2.049 & 0.040 & 0.109 & 4.837\\
\cmidrule{3-10}\nopagebreak
\hspace{1em} &  & confidence &  & -1.214 & 1.810 & -0.671 & 0.503 & -4.820 & 2.279\\
\cmidrule{3-10}\nopagebreak
\hspace{1em} &  & accuracy * confidence & * & -6.672 & 2.553 & -2.613 & 0.009 & -11.666 & -1.654\\
\cmidrule{1-10}\pagebreak[0]
\addlinespace[0.3em]
\multicolumn{10}{l}{\textbf{Faster R-CNN}}\\
\hspace{1em} & Vanishing & accuracy & * & -5.572 & 1.544 & -3.608 & 0.000 & -8.586 & -2.520\\
\cmidrule{3-10}\nopagebreak
\hspace{1em} &  & confidence & * & -6.548 & 1.557 & -4.206 & 0.000 & -9.623 & -3.513\\
\cmidrule{3-10}\nopagebreak
\hspace{1em} &  & accuracy * confidence & * & 6.505 & 2.134 & 3.047 & 0.002 & 2.327 & 10.700\\
\cmidrule{2-10}\nopagebreak
\hspace{1em} & Mislabeling & accuracy &  & -4.008 & 2.072 & -1.935 & 0.053 & -7.990 & 0.140\\
\cmidrule{3-10}\nopagebreak
\hspace{1em} &  & confidence & * & -10.366 & 2.631 & -3.940 & 0.000 & -15.562 & -5.263\\
\cmidrule{3-10}\nopagebreak
\hspace{1em} &  & accuracy * confidence & * & 8.374 & 3.358 & 2.494 & 0.013 & 1.781 & 14.920\\
\cmidrule{2-10}\nopagebreak
\hspace{1em} & Untargeted & accuracy & * & -3.045 & 1.151 & -2.646 & 0.008 & -5.305 & -0.788\\
\cmidrule{3-10}\nopagebreak
\hspace{1em} &  & confidence & * & -6.522 & 1.247 & -5.229 & 0.000 & -8.997 & -4.105\\
\cmidrule{3-10}\nopagebreak
\hspace{1em} &  & accuracy * confidence & * & 3.928 & 1.670 & 2.353 & 0.019 & 0.676 & 7.222\\
\cmidrule{1-10}\pagebreak[0]
\addlinespace[0.3em]
\multicolumn{10}{l}{\textbf{Cascade R-CNN}}\\
\hspace{1em} & Vanishing & accuracy & * & -3.474 & 1.409 & -2.466 & 0.014 & -6.223 & -0.691\\
\cmidrule{3-10}\nopagebreak
\hspace{1em} &  & confidence & * & -3.241 & 1.281 & -2.530 & 0.011 & -5.742 & -0.712\\
\cmidrule{3-10}\nopagebreak
\hspace{1em} &  & accuracy * confidence &  & 3.012 & 1.787 & 1.685 & 0.092 & -0.505 & 6.509\\
\cmidrule{2-10}\nopagebreak
\hspace{1em} & Mislabeling & accuracy &  & -2.849 & 1.600 & -1.780 & 0.075 & -5.961 & 0.326\\
\cmidrule{3-10}\nopagebreak
\hspace{1em} &  & confidence & * & -4.204 & 1.580 & -2.661 & 0.008 & -7.303 & -1.099\\
\cmidrule{3-10}\nopagebreak
\hspace{1em} &  & accuracy * confidence &  & 2.670 & 2.171 & 1.229 & 0.219 & -1.600 & 6.920\\
\cmidrule{2-10}\nopagebreak
\hspace{1em} & Untargeted & accuracy &  & -0.996 & 1.283 & -0.776 & 0.438 & -3.504 & 1.532\\
\cmidrule{3-10}\nopagebreak
\hspace{1em} &  & confidence &  & -2.287 & 1.256 & -1.821 & 0.069 & -4.759 & 0.171\\
\cmidrule{3-10}\nopagebreak
\hspace{1em} &  & accuracy * confidence &  & -1.014 & 1.751 & -0.579 & 0.562 & -4.446 & 2.423\\
\bottomrule
\end{longtable}
\endgroup{}

\begin{figure}[tb]

{\centering \includegraphics{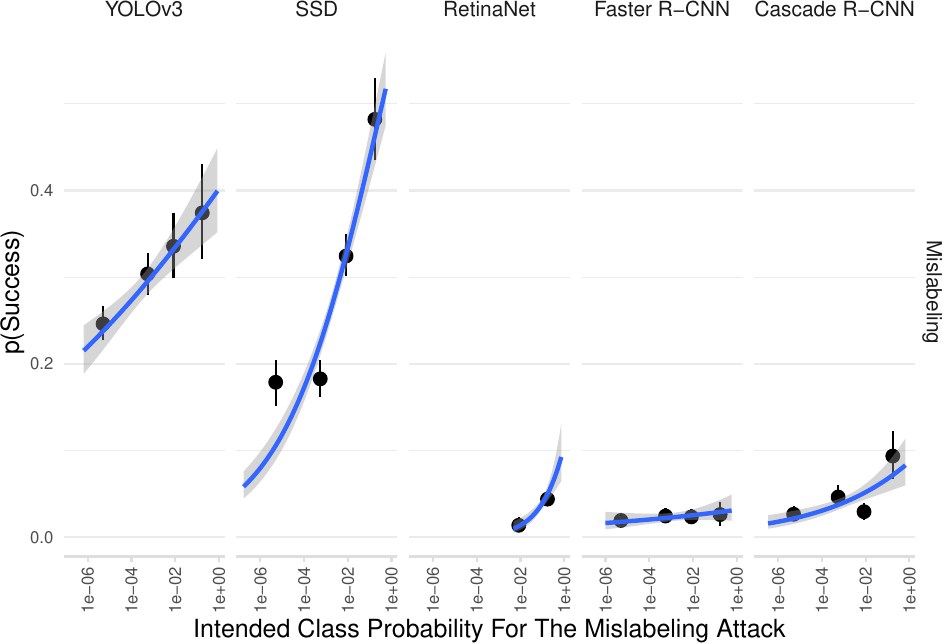} 

}

\caption{Although intended class probability seem to increase success rates for the mislabeling attack, it does not predict success rates after controlling for target class confidence, except for RetinaNet (Table \ref{tab:mislabel_conf_table}):  The binned summaries and regression trendlines graph success proportion against intended class probability in the randomized attack experiment. Bins are split into quantiles. Errors are 95\% confidence intervals}\label{fig:mislabel_conf_graph}
\end{figure}

\begingroup\fontsize{9}{11}\selectfont

\begin{longtable}[t]{llllrrrrrr}
\caption{\label{tab:mislabel_conf_table}We run a logistic model regressing success against log(intended class probability) for the mislabeling attack, with predicted class's confidence as covariate, in the randomized attack experiment. Intended class probability does not predict success rates after controlling for target class confidence, except for RetinaNet. Table headers are explained in Appendix \ref{app:tab_hdr}.}\\
\toprule
\multicolumn{2}{c}{Group} & \multicolumn{8}{c}{Regression} \\
\cmidrule(l{3pt}r{3pt}){1-2} \cmidrule(l{3pt}r{3pt}){3-10}
 & Model & term & sig & estimate & std.error & statistic & p.value & conf.low & conf.high\\
\midrule
\addlinespace[0.3em]
\multicolumn{10}{l}{\textbf{Mislabeling}}\\
\hspace{1em} & YOLOv3 & log(probability) & * & -0.202 & 0.040 & -5.028 & 0.000 & -0.281 & -0.123\\
\cmidrule{3-10}\nopagebreak
\hspace{1em} &  & confidence &  & 0.758 & 0.485 & 1.563 & 0.118 & -0.192 & 1.712\\
\cmidrule{3-10}\nopagebreak
\hspace{1em} &  & log(probability) * confidence & * & 0.363 & 0.057 & 6.337 & 0.000 & 0.251 & 0.476\\
\cmidrule{2-10}\nopagebreak
\hspace{1em} & SSD & log(probability) &  & 0.058 & 0.047 & 1.242 & 0.214 & -0.033 & 0.150\\
\cmidrule{3-10}\nopagebreak
\hspace{1em} &  & confidence &  & -0.161 & 0.429 & -0.375 & 0.707 & -1.001 & 0.682\\
\cmidrule{3-10}\nopagebreak
\hspace{1em} &  & log(probability) * confidence & * & 0.144 & 0.064 & 2.264 & 0.024 & 0.020 & 0.270\\
\cmidrule{2-10}\nopagebreak
\hspace{1em} & RetinaNet & log(probability) & * & 0.683 & 0.325 & 2.101 & 0.036 & 0.036 & 1.308\\
\cmidrule{3-10}\nopagebreak
\hspace{1em} &  & confidence & * & -8.137 & 1.846 & -4.408 & 0.000 & -11.802 & -4.567\\
\cmidrule{3-10}\nopagebreak
\hspace{1em} &  & log(probability) * confidence &  & -0.842 & 0.703 & -1.198 & 0.231 & -2.183 & 0.571\\
\cmidrule{2-10}\nopagebreak
\hspace{1em} & Faster R-CNN & log(probability) &  & 0.018 & 0.115 & 0.156 & 0.876 & -0.209 & 0.242\\
\cmidrule{3-10}\nopagebreak
\hspace{1em} &  & confidence & * & -5.405 & 1.292 & -4.183 & 0.000 & -7.955 & -2.880\\
\cmidrule{3-10}\nopagebreak
\hspace{1em} &  & log(probability) * confidence &  & -0.165 & 0.167 & -0.987 & 0.324 & -0.489 & 0.167\\
\cmidrule{2-10}\nopagebreak
\hspace{1em} & Cascade R-CNN & log(probability) &  & -0.022 & 0.095 & -0.237 & 0.813 & -0.210 & 0.162\\
\cmidrule{3-10}\nopagebreak
\hspace{1em} &  & confidence &  & -1.592 & 0.871 & -1.827 & 0.068 & -3.282 & 0.139\\
\cmidrule{3-10}\nopagebreak
\hspace{1em} &  & log(probability) * confidence &  & 0.094 & 0.124 & 0.756 & 0.450 & -0.146 & 0.340\\
\bottomrule
\end{longtable}
\endgroup{}

\begin{figure}[tb]

{\centering \includegraphics{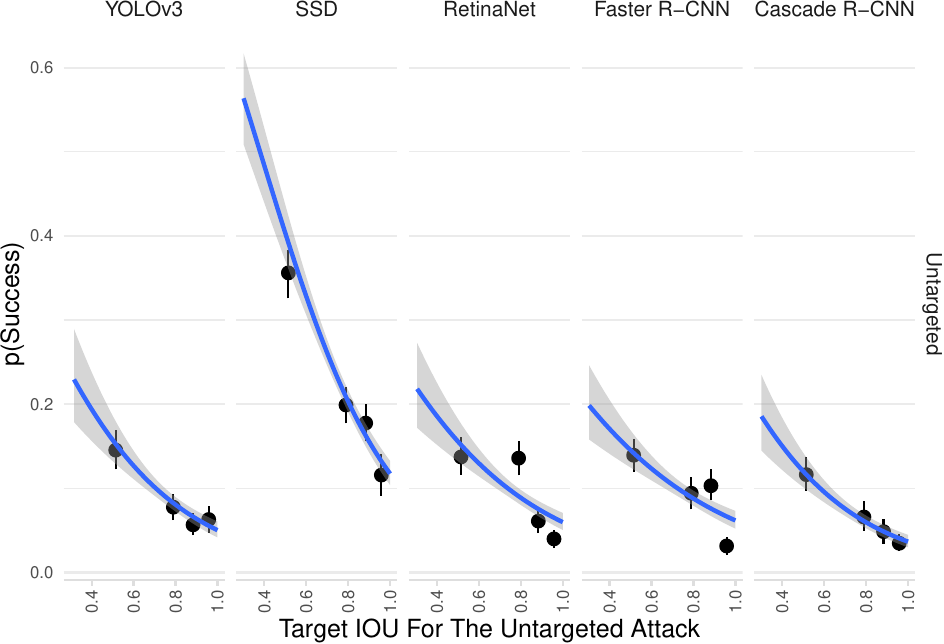} 

}

\caption{Target IOU for the untargeted attack increases success rates on all models:  The binned summaries and regression trendlines graph success proportion against target IOU for the untargeted attack in the randomized attack experiment. Bins are split into quantiles. Errors are 95\% confidence intervals}\label{fig:untarget_iou_graph}
\end{figure}

\begingroup\fontsize{9}{11}\selectfont

\begin{longtable}[t]{llllrrrrrr}
\caption{\label{tab:untarget_iou_table}We run a logistic model regressing success against target IOU for the untargeted attack in the randomized attack experiment. Target IOU for the untargeted attack increases success rates on all models. Table headers are explained in Appendix \ref{app:tab_hdr}.}\\
\toprule
\multicolumn{2}{c}{Group} & \multicolumn{8}{c}{Regression} \\
\cmidrule(l{3pt}r{3pt}){1-2} \cmidrule(l{3pt}r{3pt}){3-10}
 & Model & term & sig & estimate & std.error & statistic & p.value & conf.low & conf.high\\
\midrule
\addlinespace[0.3em]
\multicolumn{10}{l}{\textbf{Untargeted}}\\
\hspace{1em} & YOLOv3 & bbox\_iou\_eval & * & -2.526 & 0.341 & -7.417 & 0 & -3.189 & -1.853\\
\cmidrule{2-10}\nopagebreak
\hspace{1em} & SSD & bbox\_iou\_eval & * & -3.254 & 0.235 & -13.838 & 0 & -3.716 & -2.794\\
\cmidrule{2-10}\nopagebreak
\hspace{1em} & RetinaNet & bbox\_iou\_eval & * & -2.130 & 0.308 & -6.904 & 0 & -2.730 & -1.520\\
\cmidrule{2-10}\nopagebreak
\hspace{1em} & Faster R-CNN & bbox\_iou\_eval & * & -1.899 & 0.294 & -6.460 & 0 & -2.471 & -1.318\\
\cmidrule{2-10}\nopagebreak
\hspace{1em} & Cascade R-CNN & bbox\_iou\_eval & * & -2.566 & 0.318 & -8.062 & 0 & -3.187 & -1.938\\
\bottomrule
\end{longtable}
\endgroup{}

\begin{figure}[tb]

{\centering \includegraphics[width=1\linewidth]{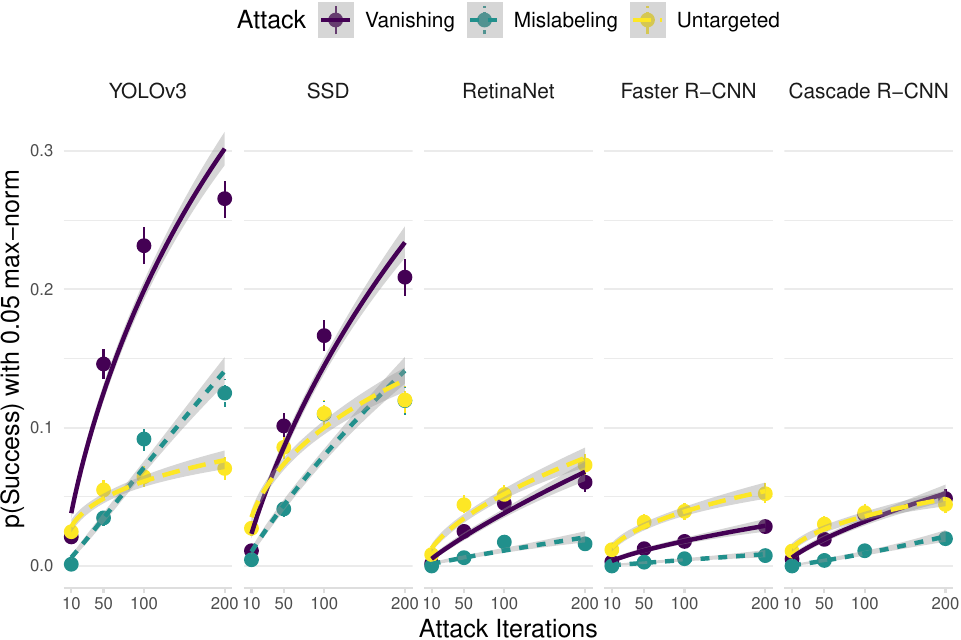} 

}

\caption{Intent obfuscating attack is feasible for all models and attacks even with 0.05 max-norm:  We conduct a randomized experiment by resampling COCO images, and within those images randomly sampling correctly predicted target and perturb objects. Then we distort the perturb objects to disrupt the target objects varying the attack iterations. The binned summaries and regression trendlines graph success proportion against attack iterations in the randomized attack experiment. Errors are 95\% confidence intervals and every point aggregates success over 4,000 images. Targeted vanishing and mislabeling attacks obtain significantly greater success on the 1-stage YOLOv3 and SSD than the 2-stage Faster R-CNN and Cascade R-CNN detectors. However, the 1-stage RetinaNet is as resilient as the 2-stage detectors. Moreover, success rates significantly increase with larger attack iterations. Significance is determined at $\alpha < 0.05$ using a Wald z-test on the logistic estimates. Full details are given in Section \ref{sec:rand_att}.}\label{fig:success_trend_graph_normed}
\end{figure}

\begin{figure*}[htb]
     \centering
     \begin{subfigure}[t]{0.35\textwidth}
         \centering
         \includegraphics[width=\textwidth]{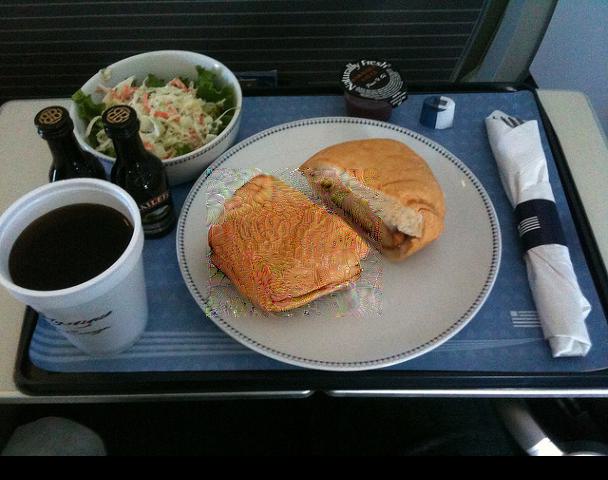}
     \end{subfigure}
     \hfill
     \begin{subfigure}[t]{0.30\textwidth}
         \centering
         \includegraphics[width=\textwidth]{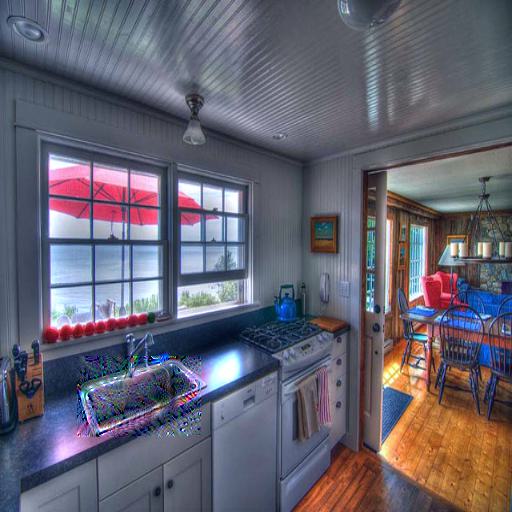}
     \end{subfigure}
     \hfill
     \begin{subfigure}[t]{0.25\textwidth}
         \centering
         \includegraphics[width=\textwidth]{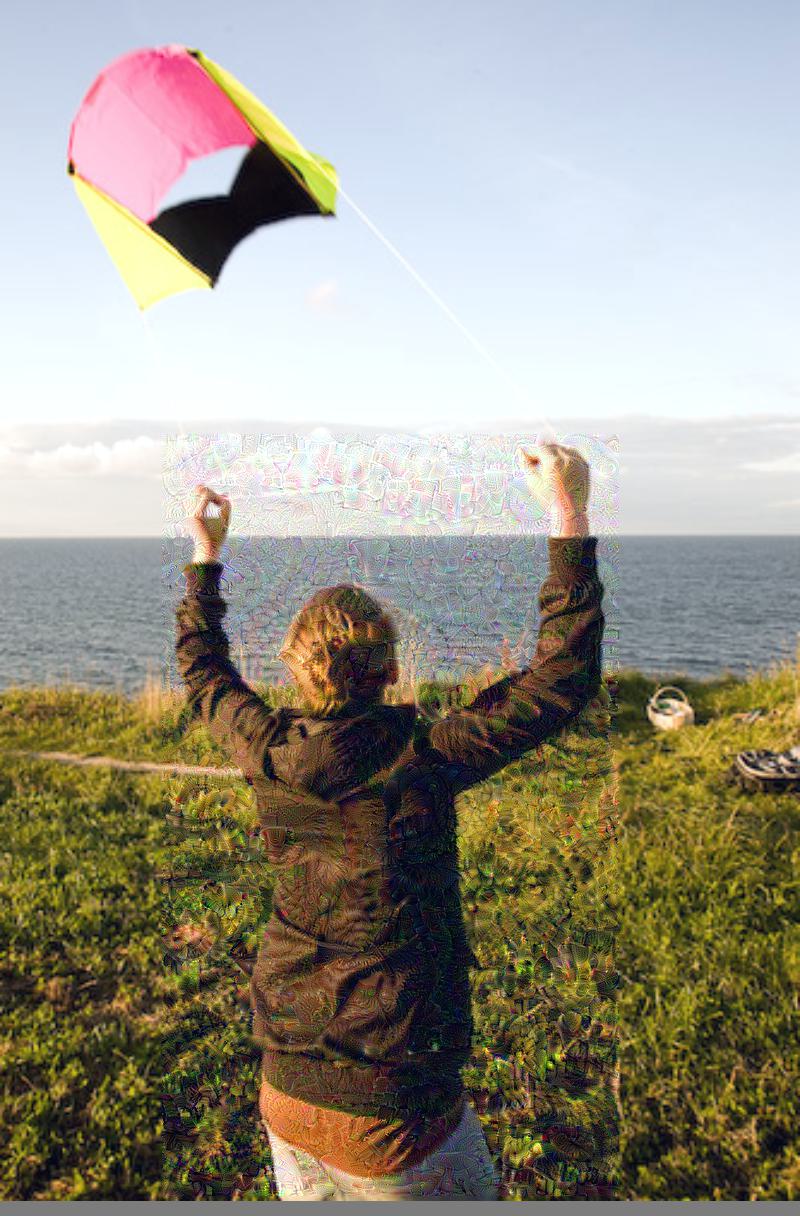}
     \end{subfigure}
        \caption{The perturbed images corresponding to the attacked examples illustrated in Figure \ref{fig:rand_img}.} 
        \label{fig:rand_pert}
\end{figure*}

\section{Deliberate Attack}

\subsection{Selecting Easier Targets}\label{app:del_per}
Since we are using a shared computing resource on an internal network, we split the attack into 2 repetitions and attacked 100 images per repetition. The images are randomly sampled without replacement within repetitions, but may repeat across repetitions. Every repetition takes approximately 30 minutes on a 32GB NVIDIA Tesla V100 GPU. 480 repetitions (5 models * 3 attacks * 2 confidences * 2 perturb-target distances * 2 bbox distances * 2 repetitions * 2 norms) take 10 V100 GPU days. More complex models (e.g. Cascade R-CNN) require more attack time than less complex models (e.g. YOLOv3).

\begin{figure}[htb]

{\centering \includegraphics[width=0.99\textwidth]{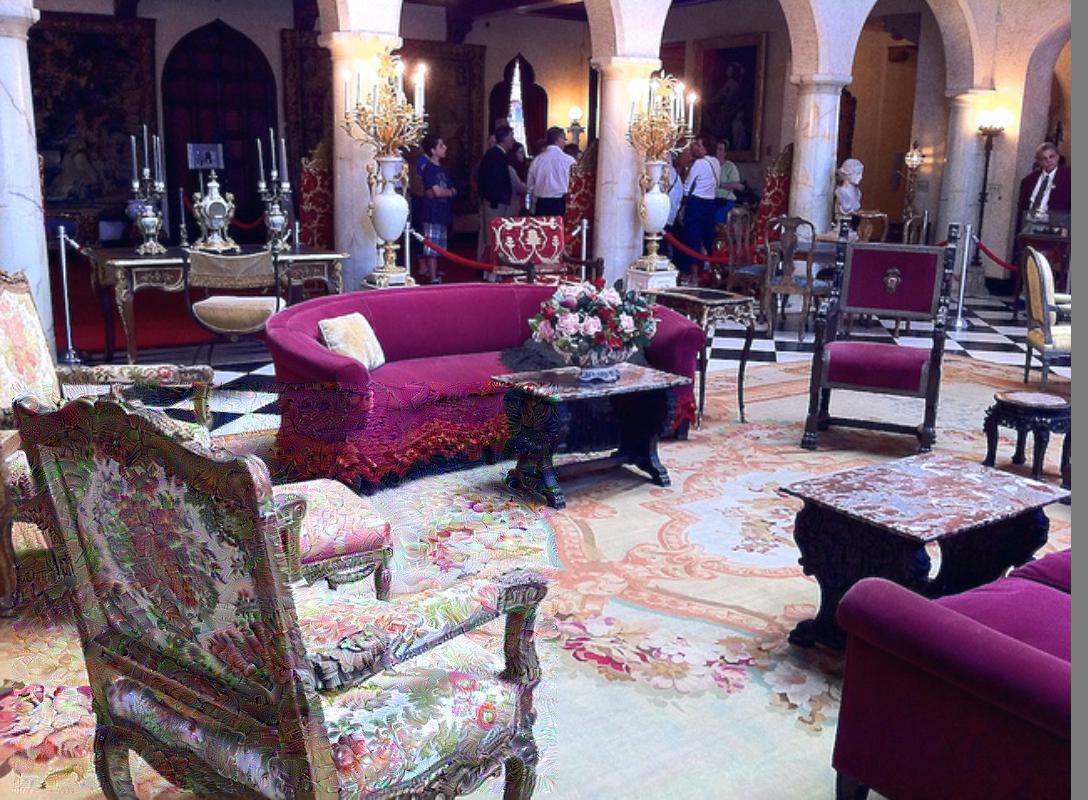} 
}

\caption{The perturbed image corresponding to the attacked example illustrated in Figure \ref{fig:del_per_exp}.}\label{fig:del_per_pert}
\end{figure}

\begingroup\fontsize{9}{11}\selectfont

\begin{longtable}[t]{llllrrrrrr}
\caption{\label{tab:num_cri_table}We run a logistic model regressing success against log(number of factors) in the randomized attack experiment. Success rates increase with the number of factors combined to select target and perturb objects for all models and attacks. Table headers are explained in Appendix \ref{app:tab_hdr}.}\\
\toprule
\multicolumn{2}{c}{Group} & \multicolumn{8}{c}{Regression} \\
\cmidrule(l{3pt}r{3pt}){1-2} \cmidrule(l{3pt}r{3pt}){3-10}
 & Attack & term & sig & estimate & std.error & statistic & p.value & conf.low & conf.high\\
\midrule
\addlinespace[0.3em]
\multicolumn{10}{l}{\textbf{YOLOv3}}\\
\hspace{1em} & Vanishing & num\_cri & * & 1.144 & 0.077 & 14.871 & 0 & 0.996 & 1.298\\
\cmidrule{2-10}\nopagebreak
\hspace{1em} & Mislabeling & num\_cri & * & 1.179 & 0.078 & 15.094 & 0 & 1.029 & 1.335\\
\cmidrule{2-10}\nopagebreak
\hspace{1em} & Untargeted & num\_cri & * & 1.007 & 0.073 & 13.700 & 0 & 0.865 & 1.153\\
\cmidrule{1-10}\pagebreak[0]
\addlinespace[0.3em]
\multicolumn{10}{l}{\textbf{SSD}}\\
\hspace{1em} & Vanishing & num\_cri & * & 0.749 & 0.065 & 11.549 & 0 & 0.624 & 0.878\\
\cmidrule{2-10}\nopagebreak
\hspace{1em} & Mislabeling & num\_cri & * & 0.684 & 0.064 & 10.752 & 0 & 0.561 & 0.810\\
\cmidrule{2-10}\nopagebreak
\hspace{1em} & Untargeted & num\_cri & * & 0.678 & 0.065 & 10.497 & 0 & 0.552 & 0.806\\
\cmidrule{1-10}\pagebreak[0]
\addlinespace[0.3em]
\multicolumn{10}{l}{\textbf{RetinaNet}}\\
\hspace{1em} & Vanishing & num\_cri & * & 0.546 & 0.086 & 6.315 & 0 & 0.378 & 0.717\\
\cmidrule{2-10}\nopagebreak
\hspace{1em} & Mislabeling & num\_cri & * & 0.586 & 0.126 & 4.657 & 0 & 0.342 & 0.836\\
\cmidrule{2-10}\nopagebreak
\hspace{1em} & Untargeted & num\_cri & * & 0.951 & 0.071 & 13.302 & 0 & 0.813 & 1.093\\
\cmidrule{1-10}\pagebreak[0]
\addlinespace[0.3em]
\multicolumn{10}{l}{\textbf{Faster R-CNN}}\\
\hspace{1em} & Vanishing & num\_cri & * & 0.558 & 0.088 & 6.319 & 0 & 0.387 & 0.733\\
\cmidrule{2-10}\nopagebreak
\hspace{1em} & Mislabeling & num\_cri & * & 0.771 & 0.107 & 7.202 & 0 & 0.564 & 0.984\\
\cmidrule{2-10}\nopagebreak
\hspace{1em} & Untargeted & num\_cri & * & 1.228 & 0.077 & 16.021 & 0 & 1.080 & 1.381\\
\cmidrule{1-10}\pagebreak[0]
\addlinespace[0.3em]
\multicolumn{10}{l}{\textbf{Cascade R-CNN}}\\
\hspace{1em} & Vanishing & num\_cri & * & 0.694 & 0.078 & 8.847 & 0 & 0.542 & 0.849\\
\cmidrule{2-10}\nopagebreak
\hspace{1em} & Mislabeling & num\_cri & * & 0.765 & 0.089 & 8.623 & 0 & 0.594 & 0.942\\
\cmidrule{2-10}\nopagebreak
\hspace{1em} & Untargeted & num\_cri & * & 0.948 & 0.075 & 12.714 & 0 & 0.804 & 1.096\\
\bottomrule
\end{longtable}
\endgroup{}

\begin{figure}[tb]

{\centering \includegraphics[width=1\linewidth]{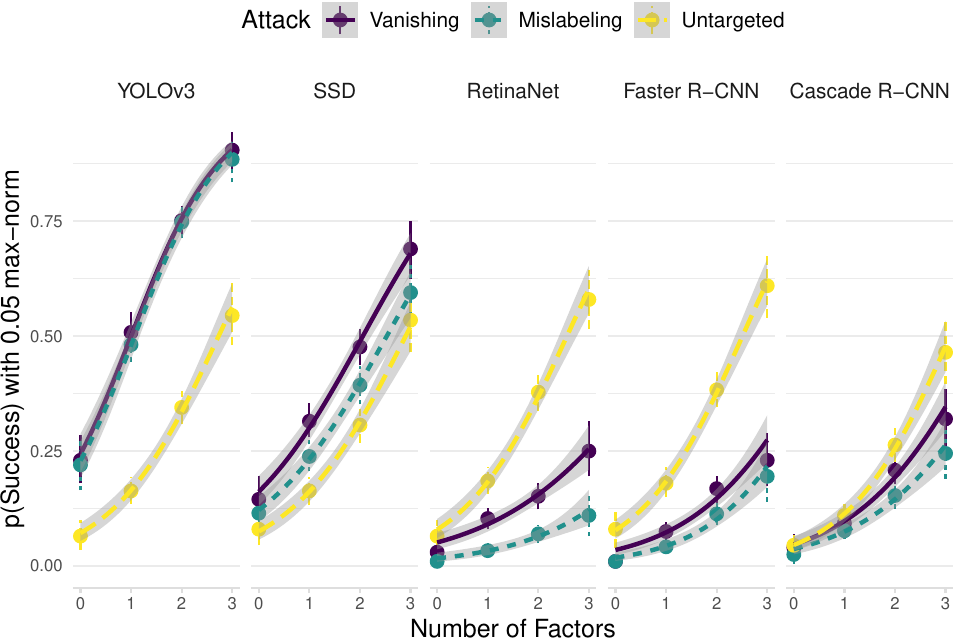} 

}

\caption{Success factors can be exploited in combination to significantly increase success rates even with 0.05 max-norm:  We sampled target and perturb objects based on three validated success factors in Table \ref{tab:results_table} by targeting objects with low predicted confidence, perturbing large objects and selecting target and perturb objects close to one another. The binned summaries and regression trendlines graph success proportion against number of factors in the deliberate attack experiment. Errors are 95\% confidence intervals and every point aggregates success over 200 images. Success rates significantly increase as the number of factors combined increases. Significance is determined at $\alpha < 0.05$ using a Wald z-test on the logistic estimates. Full details are given in Section \ref{sec:del_per}.}\label{fig:biased_trend_graph_normed}
\end{figure}

\subsection{Perturbing Arbitrary Regions}\label{app:del_arb}
Since we are using a shared computing resource on an internal network, we split the attack into 4 repetitions and attacked 50 images per repetition. The images are randomly sampled without replacement within repetitions, but may repeat across repetitions. Every repetition takes approximately 15 minutes on a 32GB NVIDIA Tesla V100 GPU. 1920 repetitions (5 models * 3 attacks * 4 perturb box lengths * 4 perturb-target distances * 4 repetitions * 2 norms) take 20 V100 GPU days. More complex models (e.g. Cascade R-CNN) require more attack time than less complex models (e.g. YOLOv3).

\begin{figure}[htb]

{\centering \includegraphics[width=0.99\textwidth]{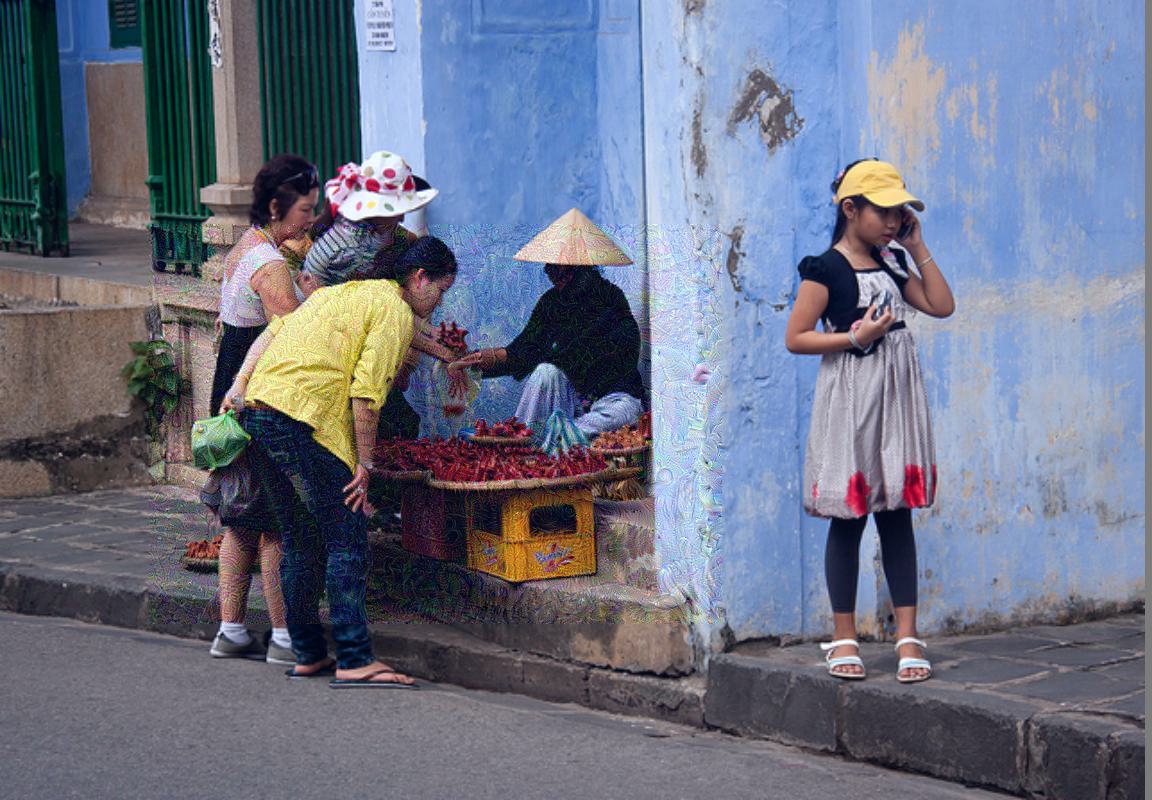} 
}

\caption{The perturbed image corresponding to the attacked example illustrated in Figure \ref{fig:del_arb_exp}.}\label{fig:del_arb_pert}
\end{figure}

\begingroup\fontsize{9}{11}\selectfont

\begin{longtable}[t]{llllrrrrrr}
\caption{\label{tab:arbitrary_trend_table}We run a logistic model regressing success against perturb-target distance and perturb box length, both relative to image width or height, in the deliberate attack experiment. Longer perturb box length or shorter perturb-target distance cause success rates to significantly increase for all model and attack combinations, except for perturb box length in untargeted attack on Cascade R-CNN. The interaction terms, even when significant, are negligibly close to 0. Table headers are explained in Appendix \ref{app:tab_hdr}.}\\
\toprule
\multicolumn{2}{c}{Group} & \multicolumn{8}{c}{Regression} \\
\cmidrule(l{3pt}r{3pt}){1-2} \cmidrule(l{3pt}r{3pt}){3-10}
 & Attack & term & sig & estimate & std.error & statistic & p.value & conf.low & conf.high\\
\midrule
\addlinespace[0.3em]
\multicolumn{10}{l}{\textbf{YOLOv3}}\\
\hspace{1em} & Vanishing & distance & * & -7.152 & 1.243 & -5.753 & 0.000 & -9.610 & -4.734\\
\cmidrule{3-10}\nopagebreak
\hspace{1em} &  & length & * & 7.648 & 0.578 & 13.235 & 0.000 & 6.543 & 8.810\\
\cmidrule{3-10}\nopagebreak
\hspace{1em} &  & distance * length & * & -12.247 & 3.877 & -3.159 & 0.002 & -19.885 & -4.676\\
\cmidrule{2-10}\nopagebreak
\hspace{1em} & Mislabeling & distance & * & -7.541 & 1.239 & -6.087 & 0.000 & -9.993 & -5.135\\
\cmidrule{3-10}\nopagebreak
\hspace{1em} &  & length & * & 6.055 & 0.442 & 13.713 & 0.000 & 5.205 & 6.937\\
\cmidrule{3-10}\nopagebreak
\hspace{1em} &  & distance * length &  & 0.465 & 3.465 & 0.134 & 0.893 & -6.299 & 7.295\\
\cmidrule{2-10}\nopagebreak
\hspace{1em} & Untargeted & distance & * & -9.464 & 1.469 & -6.441 & 0.000 & -12.392 & -6.629\\
\cmidrule{3-10}\nopagebreak
\hspace{1em} &  & length & * & 2.895 & 0.287 & 10.081 & 0.000 & 2.336 & 3.463\\
\cmidrule{3-10}\nopagebreak
\hspace{1em} &  & distance * length &  & 4.370 & 2.862 & 1.527 & 0.127 & -1.201 & 10.021\\
\cmidrule{1-10}\pagebreak[0]
\addlinespace[0.3em]
\multicolumn{10}{l}{\textbf{SSD}}\\
\hspace{1em} & Vanishing & distance & * & -9.986 & 1.267 & -7.881 & 0.000 & -12.501 & -7.532\\
\cmidrule{3-10}\nopagebreak
\hspace{1em} &  & length & * & 4.189 & 0.326 & 12.840 & 0.000 & 3.556 & 4.835\\
\cmidrule{3-10}\nopagebreak
\hspace{1em} &  & distance * length &  & -1.319 & 2.772 & -0.476 & 0.634 & -6.734 & 4.138\\
\cmidrule{2-10}\nopagebreak
\hspace{1em} & Mislabeling & distance & * & -10.593 & 1.354 & -7.826 & 0.000 & -13.284 & -7.975\\
\cmidrule{3-10}\nopagebreak
\hspace{1em} &  & length & * & 5.541 & 0.362 & 15.323 & 0.000 & 4.841 & 6.259\\
\cmidrule{3-10}\nopagebreak
\hspace{1em} &  & distance * length & * & -7.154 & 2.976 & -2.404 & 0.016 & -12.974 & -1.302\\
\cmidrule{2-10}\nopagebreak
\hspace{1em} & Untargeted & distance & * & -10.787 & 1.410 & -7.652 & 0.000 & -13.594 & -8.065\\
\cmidrule{3-10}\nopagebreak
\hspace{1em} &  & length & * & 3.497 & 0.296 & 11.810 & 0.000 & 2.921 & 4.082\\
\cmidrule{3-10}\nopagebreak
\hspace{1em} &  & distance * length &  & 1.528 & 2.835 & 0.539 & 0.590 & -3.998 & 7.119\\
\cmidrule{1-10}\pagebreak[0]
\addlinespace[0.3em]
\multicolumn{10}{l}{\textbf{RetinaNet}}\\
\hspace{1em} & Vanishing & distance & * & -17.682 & 2.722 & -6.496 & 0.000 & -23.208 & -12.539\\
\cmidrule{3-10}\nopagebreak
\hspace{1em} &  & length & * & 3.479 & 0.353 & 9.849 & 0.000 & 2.793 & 4.178\\
\cmidrule{3-10}\nopagebreak
\hspace{1em} &  & distance * length & * & -27.250 & 6.138 & -4.440 & 0.000 & -39.253 & -15.183\\
\cmidrule{2-10}\nopagebreak
\hspace{1em} & Mislabeling & distance & * & -14.139 & 3.516 & -4.022 & 0.000 & -21.420 & -7.626\\
\cmidrule{3-10}\nopagebreak
\hspace{1em} &  & length & * & 2.442 & 0.399 & 6.127 & 0.000 & 1.665 & 3.227\\
\cmidrule{3-10}\nopagebreak
\hspace{1em} &  & distance * length & * & -23.945 & 7.834 & -3.056 & 0.002 & -39.181 & -8.436\\
\cmidrule{2-10}\nopagebreak
\hspace{1em} & Untargeted & distance & * & -15.950 & 2.003 & -7.964 & 0.000 & -19.953 & -12.100\\
\cmidrule{3-10}\nopagebreak
\hspace{1em} &  & length & * & 3.483 & 0.327 & 10.664 & 0.000 & 2.850 & 4.130\\
\cmidrule{3-10}\nopagebreak
\hspace{1em} &  & distance * length & * & 24.373 & 3.645 & 6.687 & 0.000 & 17.330 & 31.623\\
\cmidrule{1-10}\pagebreak[0]
\addlinespace[0.3em]
\multicolumn{10}{l}{\textbf{Faster R-CNN}}\\
\hspace{1em} & Vanishing & distance & * & -19.538 & 3.179 & -6.146 & 0.000 & -26.021 & -13.562\\
\cmidrule{3-10}\nopagebreak
\hspace{1em} &  & length & * & 3.241 & 0.360 & 8.995 & 0.000 & 2.541 & 3.953\\
\cmidrule{3-10}\nopagebreak
\hspace{1em} &  & distance * length & * & -24.042 & 6.889 & -3.490 & 0.000 & -37.462 & -10.448\\
\cmidrule{2-10}\nopagebreak
\hspace{1em} & Mislabeling & distance & * & -18.953 & 3.679 & -5.151 & 0.000 & -26.533 & -12.110\\
\cmidrule{3-10}\nopagebreak
\hspace{1em} &  & length & * & 2.001 & 0.386 & 5.187 & 0.000 & 1.249 & 2.762\\
\cmidrule{3-10}\nopagebreak
\hspace{1em} &  & distance * length &  & -14.029 & 7.793 & -1.800 & 0.072 & -29.166 & 1.402\\
\cmidrule{2-10}\nopagebreak
\hspace{1em} & Untargeted & distance & * & -19.478 & 2.004 & -9.722 & 0.000 & -23.486 & -15.630\\
\cmidrule{3-10}\nopagebreak
\hspace{1em} &  & length & * & 3.007 & 0.310 & 9.694 & 0.000 & 2.404 & 3.620\\
\cmidrule{3-10}\nopagebreak
\hspace{1em} &  & distance * length & * & 26.412 & 3.607 & 7.322 & 0.000 & 19.439 & 33.585\\
\cmidrule{1-10}\pagebreak[0]
\addlinespace[0.3em]
\multicolumn{10}{l}{\textbf{Cascade R-CNN}}\\
\hspace{1em} & Vanishing & distance & * & -24.815 & 3.450 & -7.193 & 0.000 & -31.799 & -18.282\\
\cmidrule{3-10}\nopagebreak
\hspace{1em} &  & length & * & 4.498 & 0.410 & 10.967 & 0.000 & 3.704 & 5.312\\
\cmidrule{3-10}\nopagebreak
\hspace{1em} &  & distance * length & * & -38.766 & 7.932 & -4.887 & 0.000 & -54.349 & -23.234\\
\cmidrule{2-10}\nopagebreak
\hspace{1em} & Mislabeling & distance & * & -28.520 & 4.590 & -6.214 & 0.000 & -37.922 & -19.941\\
\cmidrule{3-10}\nopagebreak
\hspace{1em} &  & length & * & 3.122 & 0.391 & 7.978 & 0.000 & 2.362 & 3.896\\
\cmidrule{3-10}\nopagebreak
\hspace{1em} &  & distance * length & * & -20.448 & 9.401 & -2.175 & 0.030 & -38.672 & -1.816\\
\cmidrule{2-10}\nopagebreak
\hspace{1em} & Untargeted & distance & * & -34.458 & 3.088 & -11.159 & 0.000 & -40.684 & -28.577\\
\cmidrule{3-10}\nopagebreak
\hspace{1em} &  & length & * & 1.746 & 0.314 & 5.556 & 0.000 & 1.134 & 2.367\\
\cmidrule{3-10}\nopagebreak
\hspace{1em} &  & distance * length & * & 39.168 & 5.001 & 7.832 & 0.000 & 29.539 & 49.150\\
\bottomrule
\end{longtable}
\endgroup{}

\begingroup\fontsize{9}{11}\selectfont

\begin{longtable}[t]{llllrrrrrr}
\caption{\label{tab:rand_arb_compare_table}We combined the data in the randomized and deliberate attack experiments to run a logistic model regressing success against object (versus non-object), with perturb-target distance and perturb box size as covariates, both relative to image width or height. The ``object'' term codes object as 1 and non-object as 0. Perturbing an object (in the randomized attack) rather than a non-object (in the deliberate attack) significantly decreases success rates for all model and attack combinations, after controlling for perturb sizes and perturb-target distances. Table headers are explained in Appendix \ref{app:tab_hdr}.}\\
\toprule
\multicolumn{2}{c}{Group} & \multicolumn{8}{c}{Regression} \\
\cmidrule(l{3pt}r{3pt}){1-2} \cmidrule(l{3pt}r{3pt}){3-10}
 & Attack & term & sig & estimate & std.error & statistic & p.value & conf.low & conf.high\\
\midrule
\addlinespace[0.3em]
\multicolumn{10}{l}{\textbf{YOLOv3}}\\
\hspace{1em} & Vanishing & object & * & -0.537 & 0.069 & -7.786 & 0.000 & -0.673 & -0.402\\
\cmidrule{3-10}\nopagebreak
\hspace{1em} &  & distance & * & -9.619 & 0.490 & -19.631 & 0.000 & -10.594 & -8.673\\
\cmidrule{3-10}\nopagebreak
\hspace{1em} &  & size & * & 16.138 & 0.963 & 16.761 & 0.000 & 14.301 & 18.075\\
\cmidrule{3-10}\nopagebreak
\hspace{1em} &  & distance * size & * & -38.994 & 5.279 & -7.387 & 0.000 & -49.534 & -28.837\\
\cmidrule{2-10}\nopagebreak
\hspace{1em} & Mislabeling & object & * & -0.622 & 0.064 & -9.731 & 0.000 & -0.747 & -0.497\\
\cmidrule{3-10}\nopagebreak
\hspace{1em} &  & distance & * & -7.946 & 0.430 & -18.471 & 0.000 & -8.802 & -7.116\\
\cmidrule{3-10}\nopagebreak
\hspace{1em} &  & size & * & 8.275 & 0.521 & 15.875 & 0.000 & 7.275 & 9.319\\
\cmidrule{3-10}\nopagebreak
\hspace{1em} &  & distance * size &  & -5.788 & 3.262 & -1.775 & 0.076 & -12.240 & 0.551\\
\cmidrule{2-10}\nopagebreak
\hspace{1em} & Untargeted & object & * & -0.776 & 0.077 & -10.107 & 0.000 & -0.928 & -0.626\\
\cmidrule{3-10}\nopagebreak
\hspace{1em} &  & distance & * & -10.294 & 0.710 & -14.502 & 0.000 & -11.713 & -8.930\\
\cmidrule{3-10}\nopagebreak
\hspace{1em} &  & size & * & 3.025 & 0.291 & 10.388 & 0.000 & 2.457 & 3.599\\
\cmidrule{3-10}\nopagebreak
\hspace{1em} &  & distance * size & * & 10.204 & 2.615 & 3.902 & 0.000 & 5.096 & 15.352\\
\cmidrule{1-10}\pagebreak[0]
\addlinespace[0.3em]
\multicolumn{10}{l}{\textbf{SSD}}\\
\hspace{1em} & Vanishing & object & * & 0.325 & 0.064 & 5.072 & 0.000 & 0.200 & 0.451\\
\cmidrule{3-10}\nopagebreak
\hspace{1em} &  & distance & * & -12.970 & 0.533 & -24.350 & 0.000 & -14.031 & -11.943\\
\cmidrule{3-10}\nopagebreak
\hspace{1em} &  & size & * & 5.319 & 0.378 & 14.081 & 0.000 & 4.590 & 6.071\\
\cmidrule{3-10}\nopagebreak
\hspace{1em} &  & distance * size &  & 1.653 & 2.648 & 0.624 & 0.533 & -3.560 & 6.824\\
\cmidrule{2-10}\nopagebreak
\hspace{1em} & Mislabeling & object &  & -0.101 & 0.064 & -1.585 & 0.113 & -0.226 & 0.024\\
\cmidrule{3-10}\nopagebreak
\hspace{1em} &  & distance & * & -11.732 & 0.553 & -21.216 & 0.000 & -12.834 & -10.666\\
\cmidrule{3-10}\nopagebreak
\hspace{1em} &  & size & * & 6.651 & 0.403 & 16.492 & 0.000 & 5.873 & 7.454\\
\cmidrule{3-10}\nopagebreak
\hspace{1em} &  & distance * size & * & -9.854 & 2.818 & -3.497 & 0.000 & -15.407 & -4.359\\
\cmidrule{2-10}\nopagebreak
\hspace{1em} & Untargeted & object &  & 0.027 & 0.064 & 0.424 & 0.672 & -0.098 & 0.152\\
\cmidrule{3-10}\nopagebreak
\hspace{1em} &  & distance & * & -12.646 & 0.597 & -21.177 & 0.000 & -13.838 & -11.497\\
\cmidrule{3-10}\nopagebreak
\hspace{1em} &  & size & * & 3.258 & 0.291 & 11.201 & 0.000 & 2.693 & 3.834\\
\cmidrule{3-10}\nopagebreak
\hspace{1em} &  & distance * size & * & 7.145 & 2.448 & 2.919 & 0.004 & 2.344 & 11.942\\
\cmidrule{1-10}\pagebreak[0]
\addlinespace[0.3em]
\multicolumn{10}{l}{\textbf{RetinaNet}}\\
\hspace{1em} & Vanishing & object & * & -0.251 & 0.085 & -2.953 & 0.003 & -0.418 & -0.085\\
\cmidrule{3-10}\nopagebreak
\hspace{1em} &  & distance & * & -28.371 & 1.624 & -17.466 & 0.000 & -31.631 & -25.264\\
\cmidrule{3-10}\nopagebreak
\hspace{1em} &  & size & * & 3.453 & 0.360 & 9.591 & 0.000 & 2.755 & 4.167\\
\cmidrule{3-10}\nopagebreak
\hspace{1em} &  & distance * size &  & -5.791 & 5.990 & -0.967 & 0.334 & -17.676 & 5.813\\
\cmidrule{2-10}\nopagebreak
\hspace{1em} & Mislabeling & object &  & -0.164 & 0.113 & -1.447 & 0.148 & -0.388 & 0.057\\
\cmidrule{3-10}\nopagebreak
\hspace{1em} &  & distance & * & -28.622 & 2.391 & -11.973 & 0.000 & -33.480 & -24.110\\
\cmidrule{3-10}\nopagebreak
\hspace{1em} &  & size & * & 2.030 & 0.412 & 4.926 & 0.000 & 1.224 & 2.840\\
\cmidrule{3-10}\nopagebreak
\hspace{1em} &  & distance * size &  & -6.022 & 8.891 & -0.677 & 0.498 & -23.711 & 11.158\\
\cmidrule{2-10}\nopagebreak
\hspace{1em} & Untargeted & object & * & -0.403 & 0.079 & -5.130 & 0.000 & -0.558 & -0.250\\
\cmidrule{3-10}\nopagebreak
\hspace{1em} &  & distance & * & -11.268 & 0.818 & -13.768 & 0.000 & -12.910 & -9.702\\
\cmidrule{3-10}\nopagebreak
\hspace{1em} &  & size & * & 3.662 & 0.292 & 12.542 & 0.000 & 3.092 & 4.237\\
\cmidrule{3-10}\nopagebreak
\hspace{1em} &  & distance * size & * & 26.886 & 2.757 & 9.753 & 0.000 & 21.555 & 32.364\\
\cmidrule{1-10}\pagebreak[0]
\addlinespace[0.3em]
\multicolumn{10}{l}{\textbf{Faster R-CNN}}\\
\hspace{1em} & Vanishing & object & * & -0.618 & 0.104 & -5.964 & 0.000 & -0.823 & -0.416\\
\cmidrule{3-10}\nopagebreak
\hspace{1em} &  & distance & * & -27.236 & 1.889 & -14.422 & 0.000 & -31.047 & -23.643\\
\cmidrule{3-10}\nopagebreak
\hspace{1em} &  & size & * & 3.369 & 0.388 & 8.671 & 0.000 & 2.614 & 4.137\\
\cmidrule{3-10}\nopagebreak
\hspace{1em} &  & distance * size & * & -19.812 & 7.379 & -2.685 & 0.007 & -34.469 & -5.530\\
\cmidrule{2-10}\nopagebreak
\hspace{1em} & Mislabeling & object & * & -0.758 & 0.131 & -5.767 & 0.000 & -1.019 & -0.504\\
\cmidrule{3-10}\nopagebreak
\hspace{1em} &  & distance & * & -22.755 & 2.115 & -10.757 & 0.000 & -27.063 & -18.771\\
\cmidrule{3-10}\nopagebreak
\hspace{1em} &  & size & * & 2.001 & 0.412 & 4.857 & 0.000 & 1.194 & 2.810\\
\cmidrule{3-10}\nopagebreak
\hspace{1em} &  & distance * size &  & -14.270 & 8.311 & -1.717 & 0.086 & -30.831 & 1.768\\
\cmidrule{2-10}\nopagebreak
\hspace{1em} & Untargeted & object & * & -0.296 & 0.080 & -3.719 & 0.000 & -0.452 & -0.140\\
\cmidrule{3-10}\nopagebreak
\hspace{1em} &  & distance & * & -11.447 & 0.779 & -14.701 & 0.000 & -13.004 & -9.953\\
\cmidrule{3-10}\nopagebreak
\hspace{1em} &  & size & * & 3.748 & 0.304 & 12.322 & 0.000 & 3.155 & 4.347\\
\cmidrule{3-10}\nopagebreak
\hspace{1em} &  & distance * size & * & 27.445 & 2.829 & 9.703 & 0.000 & 21.965 & 33.056\\
\cmidrule{1-10}\pagebreak[0]
\addlinespace[0.3em]
\multicolumn{10}{l}{\textbf{Cascade R-CNN}}\\
\hspace{1em} & Vanishing & object & * & -0.779 & 0.097 & -7.999 & 0.000 & -0.971 & -0.589\\
\cmidrule{3-10}\nopagebreak
\hspace{1em} &  & distance & * & -29.119 & 1.854 & -15.710 & 0.000 & -32.850 & -25.584\\
\cmidrule{3-10}\nopagebreak
\hspace{1em} &  & size & * & 5.752 & 0.446 & 12.907 & 0.000 & 4.894 & 6.642\\
\cmidrule{3-10}\nopagebreak
\hspace{1em} &  & distance * size & * & -55.876 & 8.604 & -6.494 & 0.000 & -73.094 & -39.336\\
\cmidrule{2-10}\nopagebreak
\hspace{1em} & Mislabeling & object & * & -0.616 & 0.110 & -5.592 & 0.000 & -0.833 & -0.401\\
\cmidrule{3-10}\nopagebreak
\hspace{1em} &  & distance & * & -31.146 & 2.387 & -13.046 & 0.000 & -35.990 & -26.630\\
\cmidrule{3-10}\nopagebreak
\hspace{1em} &  & size & * & 3.180 & 0.381 & 8.347 & 0.000 & 2.438 & 3.933\\
\cmidrule{3-10}\nopagebreak
\hspace{1em} &  & distance * size & * & -24.457 & 9.159 & -2.670 & 0.008 & -42.647 & -6.724\\
\cmidrule{2-10}\nopagebreak
\hspace{1em} & Untargeted & object & * & -0.328 & 0.089 & -3.701 & 0.000 & -0.502 & -0.155\\
\cmidrule{3-10}\nopagebreak
\hspace{1em} &  & distance & * & -17.329 & 1.148 & -15.089 & 0.000 & -19.637 & -15.134\\
\cmidrule{3-10}\nopagebreak
\hspace{1em} &  & size & * & 2.749 & 0.298 & 9.221 & 0.000 & 2.166 & 3.335\\
\cmidrule{3-10}\nopagebreak
\hspace{1em} &  & distance * size & * & 22.929 & 3.289 & 6.972 & 0.000 & 16.523 & 29.419\\
\bottomrule
\end{longtable}
\endgroup{}

\begin{figure}[tb]

{\centering \includegraphics[width=1\linewidth]{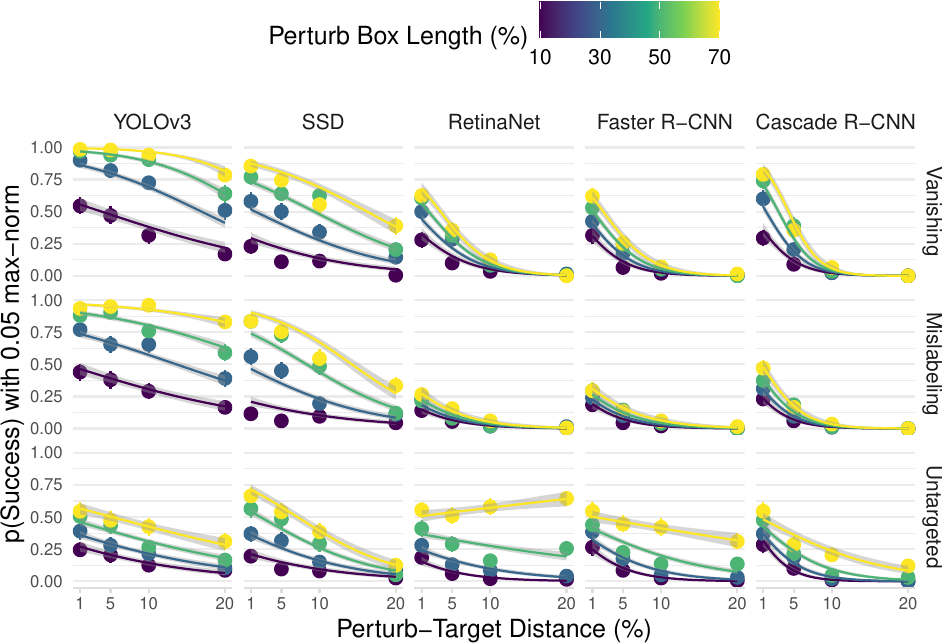} 

}

\caption{Perturbing an arbitrary region obfuscates intent with increased success for all models and attacks even with 0.05 max-norm:  We implement intent obfuscating attack by perturbing an arbitrary non-overlapping square region to disrupt a randomly selected target object at various lengths and distances. The binned summaries and regression trendlines graph success proportion against perturb-target distance and perturb box length, both relative to image width or height, in the deliberate attack experiment. Errors are 95\% confidence intervals and every point aggregates success over 200 images. The deliberate attack multiplies success as compared to the randomized attack (Figure \ref{fig:success_trend_graph}), especially at close perturb-target distance and large perturb box length. Full details are given in Section \ref{sec:del_arb}.}\label{fig:arbitrary_trend_graph_normed}
\end{figure}

\begin{figure}[htbp]
    \centering
    \includegraphics[width=\textwidth]{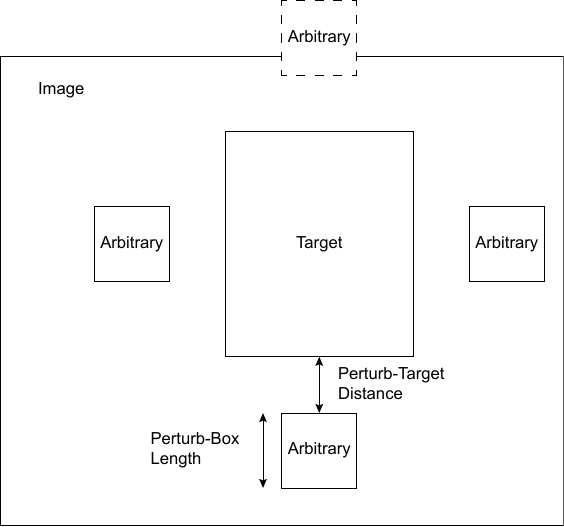}
    \caption{We randomly place a non-overlapping square perturb region region to the left, right, top, or bottom of the target object, as illustrated. The square perturb region is axes- and center-aligned to the target bounding box, and the perturb-target distance is the shortest distance between the perturb and target boundaries. We randomly sample among the eligible directions in which the perturb region is within image bounds. In the illustrated example, the top dashed region is not eligible. When all directions are not eligible, we discard the image and resample. Across model and attack combinations, we sample the same images and select the same target object and perturb direction per image to more accurately compare the success rates between combinations. In addition, if the perturb region is on the left or right, we use the image width to set the perturb box length and perturb-target distance, or else we use the image height.}
    \label{fig:del_arb_pic}
\end{figure}


\end{document}